\documentclass[aps, twocolumn, pre, notitlepage, superscriptaddress, floatfix]{revtex4-1}
\usepackage{graphicx, natbib,titlesec,amssymb, amsmath, paralist, hyperref}

\pdfoutput=1 
\def\*#1{\mathbf{#1}}
\footnotetext{A.P. and M.F. contributed equally to this work.\\ Contact: mfreilich@ucsd.edu}

\begin{document}

\title{Oceanic frontal divergence alters phytoplankton competition and distribution}
\author{Abigail Plummer$^*$}
\affiliation{Department of Physics, Harvard University, Cambridge, MA}
\affiliation{Princeton Center for Complex Materials, Princeton University, Princeton, NJ}

\author{Mara Freilich$^*$}
\affiliation{MIT-Woods Hole Oceanographic Institution Joint Program in Oceanography and Applied Ocean Science, Woods Hole, MA}
\affiliation{Scripps Institution of Oceanography, University of California San Diego, San Diego, CA}

\author{Roberto Benzi}
\affiliation{Department of Physics and Istituto Nazionale di Fisica Nucleare, University of Rome Tor Vergata, Rome, Italy}

\author{Chang Jae Choi}
\affiliation{GEOMAR Helmholtz Centre for Ocean Research, Kiel, Germany}
\affiliation{University of California Santa Cruz, Santa Cruz, CA, USA}

\author{Lisa Sudek}
\affiliation{University of California Santa Cruz, Santa Cruz, CA, USA}

\author{Alexandra Z. Worden}
\affiliation{GEOMAR Helmholtz Centre for Ocean Research, Kiel, Germany}
\affiliation{University of California Santa Cruz, Santa Cruz, CA, USA}
\affiliation{Marine Biological Laboratory, Woods Hole, MA, USA}

\author{Federico Toschi}
\affiliation{Department of Applied Physics, Eindhoven University of Technology, 5600 MB Eindhoven, The Netherlands}
\affiliation{Istituto per le Applicazioni del Calcolo, Consiglio Nazionale delle Ricerche, 00185 Rome, Italy}

\author{Amala Mahadevan}
\affiliation{Department of Physical Oceanography, Woods Hole Oceanographic Institution, Woods Hole, MA 02543}

\begin{abstract}
Ecological interactions among phytoplankton occur in a moving fluid environment. Oceanic flows can modulate the competition and coexistence between phytoplankton populations, which in turn can affect ecosystem function and biogeochemical cycling. We explore the impact of submesoscale velocity gradients on phytoplankton ecology using observations, simulations, and theory. Observations reveal that the relative abundance of {\em Synechoccocus} oligotypes varies on 1--10~km scales at an ocean front with submesoscale velocity gradients at the same scale. Simulations in realistic flow fields demonstrate that regions of divergence in the horizontal flow field can substantially modify ecological competition and dispersal on timescales of hours to days. Regions of positive (negative) divergence provide an advantage (disadvantage) to local populations, resulting in up to $\sim~20\%$ variation in community composition in our model. We propose that submesoscale divergence is a plausible contributor to observed taxonomic variability at oceanic fronts, and can lead to regional variability in community composition.
\end{abstract}

\date{\today}

\maketitle
\thispagestyle{empty}

Phytoplankton form the base of the marine food web and mediate ocean uptake of carbon and oxygen. Therefore, phytoplankton abundance and diversity are important determinants of the health of the ocean, and quantitative theories of marine ecology are required to make precise predictions about our changing climate \cite{weber2012oceanic,vallina2014global}. Although many factors such as nutrients, sunlight, and temperature are well-established drivers of community composition, they are often not sufficient to explain the observed spatial patterns of variability in phytoplankton community structure \cite{clayton2017co,ramond2021phytoplankton} or high levels of plankton diversity \cite{hutchinson1961paradox,biller2015prochlorococcus}.

Some of the unaccounted-for variability is likely introduced by the physical flows transporting planktonic organisms \cite{speirs2001population, lutscher2007spatial, wares2008drift}. Oceanic flow fields shape marine ecosystems because plankton swim much slower than the speed of ocean currents \cite{mahadevan2002biogeochemical,mahadevan2016impact,mcgillicuddy2016mechanisms, herrerias2019motion}.  At the mesoscale (10~km and larger), ocean currents are primarily horizontal---the typical magnitude of horizontal velocity is $\mathcal{O}(10^{-1})$~$\mathrm{m s}^{-1}$ while the typical magnitude of vertical velocity is $\mathcal{O}(10^{-5})$~$\mathrm{m s}^{-1}$. Lateral stirring and mixing by ocean currents disperses organisms, which can increase diversity \cite{d2010fluid,clayton2013dispersal,barton2010patterns,kashtan2014single, villamartinprotists, karolyi2000chaotic}. However, at the submesoscale (1--10~km spatial scales), there are a number of processes, including waves and frontal dynamics, that can increase the magnitude of the vertical velocity to $\mathcal{O}(10^{-3})$~$\mathrm{m s}^{-1}$ and increase the horizontal velocity divergence to the same order as the planetary vorticity (the Coriolis frequency) \cite{d2018ocean,barkan2019role}. Due to both the lack of observations and lack of theoretical understanding of the range of effects submesoscales may have on ocean biology, these types of dynamics have been largely overlooked. Submesoscale dynamics are particularly challenging to study because of their small size and fast temporal evolution, which often eludes modern observational technology and challenges computational limitations of ocean models. 

Despite recent progress in the field of submesoscale dynamics \cite{mcwilliams2016submesoscale}, the ways that these dynamics influence ecological interactions are only beginning to be understood. For example, submesoscale flows can lead to higher productivity and alter community composition on timescales of days by facilitating the exchange of nutrients and organisms between the dark ocean interior and the sunlit surface layer \cite{levy2018role,uchida2020vertical,kessouri2020submesoscale,freilich2021diversity}. The increased horizontal velocity divergence displayed by submesoscale flows may be especially relevant to phytoplankton ecology, as recent theoretical work in population genetics has shown that even weak horizontal velocity divergence can affect long-term competition outcomes between organisms \cite{plummer, guccione2019discrete}. 


In this report, we investigate the ecological significance of submesoscale velocity divergence. We focus on competition between phytoplankton subpopulations with similar growth characteristics in both our observations and our model in order to isolate and quantify the effects of the physical flow field. We begin by presenting observational evidence of unexplained variations in phytoplankton community composition at a front with regions of divergence on the edge of the eastern Alborán Gyre in the Western Mediterranean Sea. This is a regularly reoccurring front between the Atlantic and Mediterranean water masses \cite{tintore1988study,heburn1990variations}. We then discuss in more detail how a phytoplankton population in a stratified environment may experience a flow field with a nonzero divergence, generating an effective compressibility. We demonstrate the biological relevance of regions of divergence with simulations of a two-dimensional model for competition between two plankton populations in a realistic oceanographic flow field initialized with observational data. Finally, we extend and generalize the theoretical model presented in \citet{plummer} to two dimensions and short timescales to explain behavior observed in simulations.  We conclude by discussing implications for future simulations and observations of ocean ecology. 


\section{Methods}
\subsection{Observations}
Samples were collected from the sea surface at 5~min intervals using an oceanographic bucket while the ship was underway at 8~knots to obtain approximately 1~km lateral sampling resolution while crossing a front nearly perpendicularly. A total of 16 samples were collected this way on May 30, 2018 from 20:00 UTC to 22:45pm UTC (22:00 May 30 to 00:45 May 31 local time). DNA samples were obtained by filtering 500~ml seawater through 47~mm 0.2~$\mu$m pore size polyethersulfone membrane filters (Supor 200, Pall Gelman). Filters were placed into sterile cryovials, flash-frozen in liquid nitrogen where they were stored for the remainder of the research cruise. After the cruise, samples were stored at -80$^\circ$C until analysis. Sample DNA was extracted with a DNeasy Plant Kit (Qiagen), with a modification including a bead beating step \cite{demir2011global}. DNA was amplified using the primers 27FB (5$^\prime$-AGRGTTYGATYMTGGCTCAG-3$^\prime$) and 338RPL (5$^\prime$-GCWGCCWCCCGTAGGWGT-3$^\prime$) as in \cite{vergin2013high,sudek2015cyanobacterial} targeting the V1-V2 hypervariable region of the 16S rRNA gene with Illumina adapters. PCR reactions contained 25 ng of template, 5 $\mu$l of 10$\times$ buffer, 1 U of HiFi-Taq, 1.6 mM MgSO4 (Thermo Fisher) and 0.2 $\mu$M of each primer. The PCR cycling parameters were 94$^\circ$C for 2~min; 30$\times$94$^\circ$C for 15~s, 55$^\circ$C for 30~s, 68$^\circ$C for 1~min, and a final elongation at 68$^\circ$C for 7~min. Paired-end library sequencing (2 $\times$ 300bp) was performed using the Illumina MiSeq platform (Illumina). Sequences were demultiplexed and assigned to samples using CASAVA (Illumina). A 10 bp running window was utilized to trim low-quality sequence ends at a Phred quality (Q) of 25 using Sickle 1.33 \cite{joshi2011sickle}. Paired-end reads were merged using USEARCH v10.0.240 \cite{edgar2015error} when reads had a $\geq$ 50 bp overlap with maximum 5\% mismatch. The merged reads were then filtered to remove reads with maximum error rate $>$ 0.001 or shorter than 200 bp. Only sequences with exact match to both primers were kept and primer sequences were trimmed using Cutadapt v.1.13 \cite{martin2011cutadapt}. Cyanobacterial amplicons were initially parsed using the phylogenetic pipeline in PhyloAssigner v.6.166 \cite{vergin2013high} and then further classified using fine-scale cyanobacterial reference alignment and tree \cite{sudek2015cyanobacterial} according to protocols outlined in \cite{choi2017newly}. Oligotyping was then performed on aligned and trimmed samples in Qiime using oligotyping pipeline version 3.1 on 108,088 reads classified as \textit{Synechococcus IV} specifying 4 components. This resulted in 16 oligotypes which represent 99.88\% of all reads with a purity score of 1.0 \cite{eren2013oligotyping}. In Fig. \ref{fig:observations}, percent difference = $\frac{f_i-f_0}{f_0} \times 100$ where $f_i$ is the abundance of the ATTT oligotype relative to all of the sequences identified as {\em Synechococcus} IV in sample $i$ and $f_0$ is the same quantity in the reference sample. 

Samples for quantifying the cyanobacteria abundance were preserved with EM grade 25\% Glutaraldehyde (10~µl per 1 ml seawater). Samples were placed in sterile cryovials and flash frozen in liquid nitrogen and then stored in liquid nitrogen for the remainder of the research cruise after which they were stored at -80$^\circ$C until analysis. Samples were analyzed using a BD Influx flow cytometer equipped with a 488~nm laser. Calibration beds were added to each sample immediately before analysis (0.75~$\mu$m yellow-green, Polysciences, Inc and 1.0-1.4 $\mu$m ultrarainbow, Spherotech). Each sample was run for 8~min at 25~$\mu l$~min$^{-1}$ after a pre-run of 2~min. Forward angle light scatter (FALS), side scatter, and autofluorescence at 692/20~nm, 572/13~nm, and 520/25~nm were recorded, with data collection triggered by FALS. {\em Synechococcus} cells were classified using red and orange autoflourescence and FALS. 

Samples to be analyzed for nitrate were frozen at -20$^\circ$C and analyzed to determined the concentrations of nitrate, nitrite, silicate, and phosphate with a nutrient autoanalyzer (Skalar SAN++ System) at the Institute for Marine Sciences of Andalusia (ICMAN-CSIC). The limits of detection are nitrate (0.13±0.1), nitrite (0.025±0.005), phosphate (0.036±0.004), silicate (0.091±0.042).

Simultaneous with the biological sampling, we measured the depth structure of temperature and salinity with an EcoCTD towed profiler \cite{dever2020ecoctd}, the surface temperature and salinity with a thermosalinograph (SBE90) on the ship seawater intake, and the water velocity with a vessel-mounted Acoustic Doppler Current Profiler (150 kHz ADCP).

\subsection{Ocean model}
To construct realistic oceanographic flow fields, we initialize the Process Study Ocean Model (PSOM) \cite{mahadevan1996nonhydrostatic,mahadevan1996nonhydrostatic1} in thermal wind balance with density sections derived from observations of the Almer{\' i}a-Oran front sampled by a glider at 1~km horizontal resolution during the July 2017 IRENE research cruise. The boundaries of the data are interpolated to form an idealized domain that is 128~km by 206~km by 1~km in extent. The model horizontal resolution is 500~m, except near the closed north and south walls where the cell length increases linearly to 2~km. The model is evolved with a horizontal diffusivity of 1 m$^2$s$^{-1}$ and a vertical diffusivity of $10^{-5}$ m$^2$s$^{-1}$. 

The flow fields develop meanders and smaller scale divergent features, which are mostly localized at the front.  We wait until the total kinetic energy of the system has reached a steady state and record the 3D velocity fields. A snapshot of a 3D simulation can be found in \cite{freilich2021coherent}. This flow field has hydrographic and velocity gradient structure that is statistically similar to the summer season. This model summer flow field has a 5~m deep mixed layer that is lighter than any interior density surface, effectively isolating the surface from the interior. To construct the second initial condition, we deepen the mixed layer by cooling the surface and recomputing the surface density profile using convective adjustment until the maximum mixed layer depth is 70~m. This process leads to a flow field characteristic of the winter season. The winter model has a more active surface-enhanced submesoscale flow field, which results in smaller scale features in the velocity gradients \cite{freilich2021coherent}. The model is periodic in the east-west direction (parallel to the front) and has closed walls in the north and south. 

To perform simulations that couple the flow fields with the biological variables, we select 2D slices from the 3D PSOM fields. We use the surface layer in the summer model, and a slice at a depth of 52 m in the winter model, which is near the base of the mixed layer. See SI Secs. A and D for more information on the flow fields and further discussion of the validity of the constant depth approximation. 

We simulate the evolution of up to 1056 ($33 \times 32$) initial conditions centered at different locations in the domain over a 24-hour period, all experiencing the same velocity field (Fig. \ref{maps}). We simulate the population concentrations offline by stepping Eqs. \ref{ca} and \ref{cb} forward with a second-order Adams-Bashforth scheme and linearly interpolating the flow fields in time from model snapshots saved every 3 hours. The spatial derivatives in the diffusion operator are discretized using a central second-order finite-difference method. 

\section{Observations}\label{observations}
Here we present high spatial resolution observations of a chlorophyll filament in a region of velocity convergence in the Western Mediterranean Sea. These observations establish that measurable genetic variations are present at the submesoscale. This dataset is unprecedented in its combined biological and physical resolution with horizontal resolution of 1~km and the use of approaches for resolving diversity and taxonomy at high phylogenetic resolution.

\begin{figure*}[htp]
    \centering
    \includegraphics[width = 0.95\textwidth]{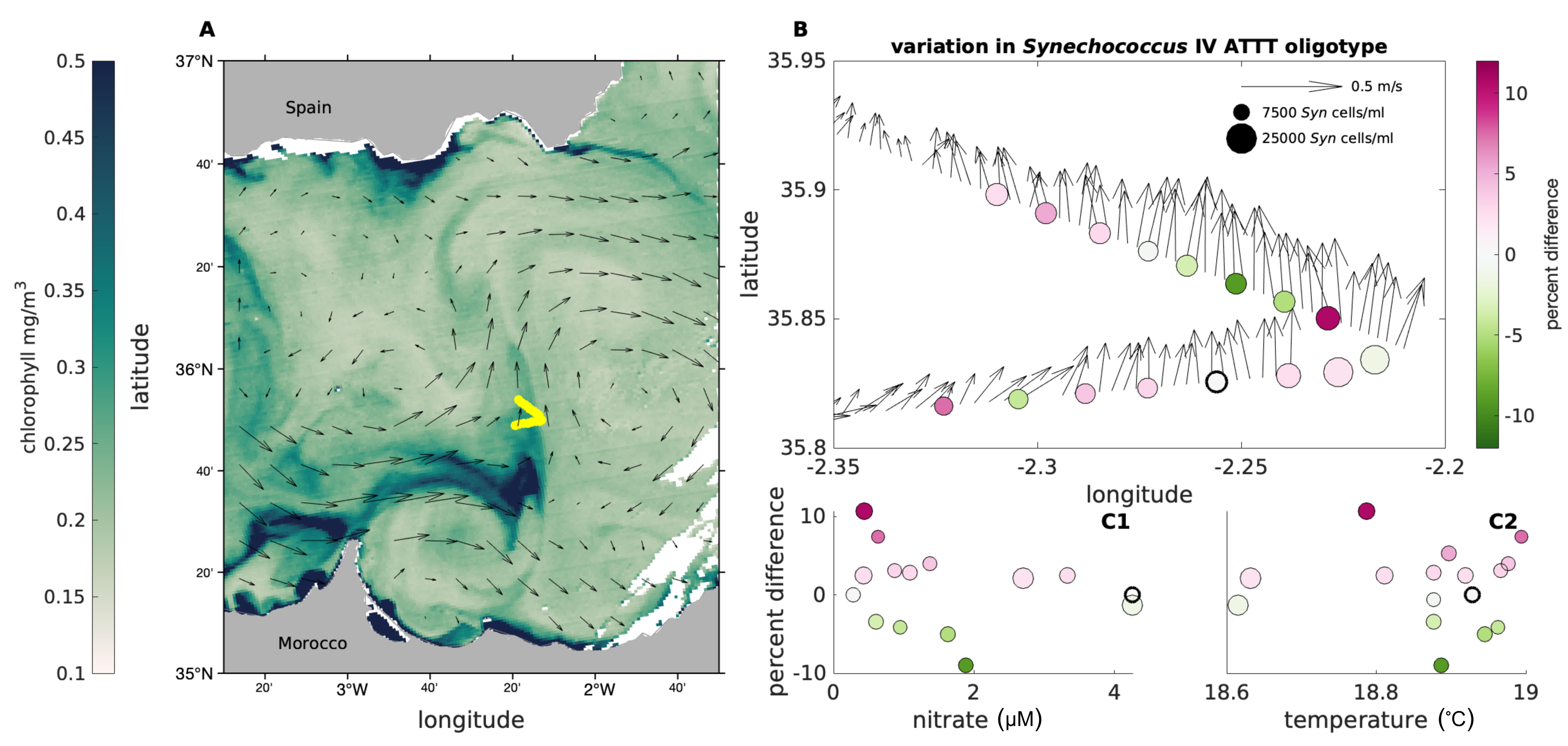}
    \caption{Relative abundances of {\em Synechococcus} IV oligotypes show no correlation with temperature or nitrate concentration across a front in the Mediterranean Sea with regions of strong divergence. (A) The yellow line (``V'' shape) shows where water was sampled across the surface layer of a chlorophyll filament in the Western Mediterranean Sea . The background shading shows chlorophyll concentration measured by satellite (MODIS) with geostrophic velocity vectors from the AVISO satellite product on May 30, 2018. (B) Spatial variation in community composition and velocity. Plotted along the ship's track shown in (A), we show the concentration of {\em Synechococcus} cells in surface waters (circle size) and the percent difference of the relative abundance of the ATTT oligotype of {\em Synechococcus} ecotype IV relative to the sample with the bold outline, based on V1-V2 16S amplicon relative abundance data (colors). The arrows indicate the velocity of the flow at the shallowest depth measured by the vessel-mounted ADCP (16~m). (C) Variability in community composition is not correlated with environmental factors. Percent difference in the relative abundance of the ATTT oligotype is plotted against nitrate concentration (C1) and temperature (C2). The colors correspond to the percent differences given for each point in (B).}
    \label{fig:observations}
\end{figure*}

The velocity convergence occurs at a front that is generated by the confluence of water from the Atlantic Ocean with the warmer and saltier water of the Mediterranean Sea \cite{tintore1988study}. This confluence leads to a fast-flowing current at the boundary of the two water masses (Figure \ref{fig:observations}A). Instabilities develop at the 100~km scale of the front and at the submesoscale as the water masses attempt to vertically stratify, with the Mediterranean water sinking below the Atlantic water, leading to increased vertical velocity, relative vorticity, and divergence. For more details, see Supporting Information (SI) Sec. A. 

We observe convergence of the velocity on the sampling track with a maximum value of $\sim 2.5 \times 10^{-4}$~s$^{-1}$ calculated along the ship track (Fig. \ref{fig:observations}B). The variability in velocity gradients on kilometer scales (the along track velocity switches from divergent to convergent on 4~km scale, SI Fig. S5) and velocity gradients of the same order as the Coriolis frequency indicates the prevalence of surface-intensified submesoscale dynamics. During this sampling, surface divergence of up to $1.3 \times 10^{-4}$~s$^{-1}$ was observed with drifters in this region \cite{tarry2021frontal}. The observed surface mixed layer is approximately 10~m deep and water parcels would not be adiabatically far from the surface mixed layer in this location, suggesting that vertical nutrient fluxes do not explain the observed population distributions. The vertical motion associated with the observed divergence likely results in perturbations of the depth of the mixed layer. In these observations, even as populations may move tens of kilometers in the horizontal, they are restricted to move only a few meters in the vertical due to the density stratification that restricts adiabatic exchange of water masses between the surface and interior \cite{tarry2021frontal}.


In the region under study we found that {\em Synechococcus} was the most abundant phytoplankter with 9,080--22,700 cells ml$^{-1}$ compared with just 630--2,000 cells ml$^{-1}$ {\em Prochlorococcus} and 2,200--5,500 cells ml$^{-1}$ eukaryotes (SI Fig. S8). 90\% of the {\em Synechococcus} cells were comprised of {\em Synechococcus} IV (SI Fig. S7), an ecotype previously observed in coastal and relatively cool waters \cite{zwirglmaier2008global}.

We further quantified microdiversity using oligotyping. An oligotype is a population defined based on subtle variations in nucleotide sequences \cite{eren2013oligotyping}. 
All oligotypes within the {\em Synechococcus} IV ecotype likely have similar gross growth characteristics \cite{pittera2014connecting,kashtan2014single,six2021marine}, although they may vary in their ecological function and food web interactions, including differential impacts from viruses \cite{jaspers2004ecological,berry2017oligotypes}. {\em Synechococcus} ecotypes in particular have been observed to encompass groups that have substantial ecological variation \cite{farrant2016delineating}.  In our observations, we detected a total of 16 {\em Synechococcus} IV oligotypes using V1-V2 16S rRNA gene amplicon sequences.

We find approximately 10\% variation in the abundance of the most abundant {\em Synechococcus} IV oligotype in our survey (oligotype ATTT) relative to the abundance of all other observed  {\em Synechococcus} IV oligotypes on the scale of the front.  The relative abundance of the ATTT oligotype does not correlate with nutrients or temperature (Figs. \ref{fig:observations} C1, C2). The observed variations in relative abundance did not necessarily originate locally. We can use satellite measurements of the geostrophic velocity to estimate where our samples were ten days prior to our observations---this calculation is shown in SI Fig. S11, and reveals no obvious correlation between mesoscale stirring and submesoscale patterns of genetic variation.  However, we were not able to make any Lagrangian measurements of diversity, such as following a single water parcel as it transited a region of divergence. Therefore, we treat the distribution of oligotypes as a single high-resolution snapshot of a community that exists in a region with strong and variable divergence of the horizontal flow field.

We conclude that the distribution of {\em Synechococcus} IV oligotypes in our observations cannot be understood in terms of the measured abiotic factors alone.  We next query whether the observed submesoscale divergence could have significantly impacted the distribution of oligotypes. However, we emphasize that the sampling and this analysis does not control for a number of other mechanisms for generating genetic variability, including variable predation or alterations in host-viral encounter rates.

\section{Model}\label{biologicalmodel}
In order to assess whether regions of divergence could have contributed to the observed genetic variability, we develop a simplified model that couples population dynamics in two dimensions to a time-varying flow field for a 24-hour period.  While the observations inform and motivate the model assumptions, due to the complexity of this system and limitations on data collection, we do not attempt to model all aspects of the observed system. Instead, by using a general, coarse-grained biological model, our study reveals broadly applicable principles driving ecological changes in frontal regions.

\subsection{Biophysical assumption: effectively 2D populations}
During the 24-hour period examined here, we assume that phytoplankton populations are restricted to live in a particular depth range within the three-dimensional flow field. This assumption is a reasonable approximation for several scenarios. The most intuitive scenario is positively buoyant organisms that cannot be subducted away from the sea surface \cite{taylor2020influence,moore1996buoyancy}. Alternately, some organisms are restricted to live in narrow subsurface depth ranges due to motility and light dependent growth and predation \cite{benoit2009edge}. The size and buoyancy characteristics of phytoplankton populations affect their depth ranges due to the impact of viscosity and physiological characteristics such as gas vesicles \cite{litchman2008trait}. In our observations, it may be reasonable to model populations as restricted to a particular depth range because there is a strong depth partitioning of the community structure with reduced abundance of \textit{Synechococcus} IV below the shallow mixed layer (SI Fig. S10).  Of course, there are also many other important scenarios in which the assumption of a population restricted to a surface does not hold \cite{freilich2021diversity}.


As discussed extensively in past work \cite{larkin2010time, boffetta2004lagrangian, de2015clustering, perlekar2010population, perlekar2013cumulative, pigolotti2012population, pigolotti2013growth, plummer, guccione2019discrete, benzi2022spatial}, the insight that some populations are restricted to remain close to a particular depth surface is consequential because such populations can experience a velocity field with nonzero divergence. For a concrete example, we again consider positively buoyant organisms at the sea surface experiencing an upwelling---organisms are spread apart by an effectively compressible flow. This argument can be extended to organisms that experience a force confining them to a sub-surface depth \cite{de2015clustering}. Additionally, while we simulate a depth-restricted population for this study, the results can also be applied to populations restricted to density surfaces. Organisms may regulate their buoyancy to stay near a particular density surface \cite{boyd2002impact}. In the models used here, the distribution of divergence for the along-isopycnal flows is similar to the distribution of divergence for flows on depth surfaces, with slightly weaker extreme values (SI Sec. D).



Applying this assumption of a depth-restricted community, we consider two-dimensional populations experiencing compressible flows in our modeling. Exploring this simplified model allows us to isolate and quantify the impact of divergence on ecological competition.


\subsection{Model flow fields}
Two models are used, one with a shallow mixed layer, as in the observations from May (called the ``summer'' model), and one in which the mixed layer has been deepened to generate surface-enhanced submesoscale dynamics (called the ``winter'' model). In the winter model the deep reaching front outcrops at the surface but in the summer model the surface layer is stratified and the density front does not outcrop. Using 24-hour periods from two different model runs allows us to examine a wider range of oceanographic conditions. A 24-hour period is long enough that we can observe population growth given the generation time used in simulations. 

\subsection{Biological model}
We use a general biological model that can serve as a starting point for understanding a wide range of competition scenarios \cite{murray}. Specifically, we consider two populations, $A$ and $B$, that compete with one another while being passively advected. We use the term ``population'' to refer to a group of organisms that shares common niche, competition, and growth characteristics (e.g. a species, ecotype, amplicon sequence variant, or oligotype). In terms of the observations of variations in the relative abundance of a {\em Synechococcus} IV oligotype reported in the previous section, population A would be the ATTT oligotype, the most abundant oligotype observed in the transect (approximately 10\% of all \textit{Synechococcus} IV cells), and population B would be all other oligotypes combined (approximately 90\% of all \textit{Synechococcus} IV cells). 


The population dynamics are modeled as a reaction-advection-diffusion system with logistic growth \cite{murray, neufeld, tel2005chemical, benzi2009fisher}. The coupled partial differential equations that describe this system can, for example, be derived by coarse-graining agent-based birth and death processes \cite{pigolotti2012population} and neglecting the noise terms due to the large population sizes (note that $N \gtrsim 10^4$ cells per ml in observations). These equations are
\begin{align}
    \frac{\partial c_{A}}{\partial t}+\nabla \cdot\left(\mathbf{u} c_{A}\right)&=D \nabla^{2} c_{A}+\mu c_{A}\left(1-c_{A}-c_{B}\right)+s \mu c_{A} c_{B},\label{ca}\\
    \frac{\partial c_{B}}{\partial t}+\nabla \cdot\left(\mathbf{u} c_{B}\right)&=D \nabla^{2} c_{B}+\mu c_{B}\left(1-c_{A}-c_{B}\right)-s \mu c_{A} c_{B}. \label{cb}
\end{align}
Here, $c_A(\*x,t)$ and $c_B(\*x,t)$ describe the concentration of the population at position $\*x$ as a fraction of the local carrying capacity of the respective population in the absence of competition and advection (i.e., when $\partial c_A/\partial t =0$ with $\*u=0$ and $c_B=0$). We note that the total concentration, $c_A+c_B$, is not required to be constant \cite{pigolotti2012population}. The diffusivity, $D$ is assumed to have the same value as the carrier fluid and is $ 1 \text{ m}^2\text{s}^{-1}$, unless otherwise noted, $\*u(\*x, t)$ is a two-dimensional velocity field with $\nabla \cdot \*u \neq 0$, and $\mu$ is the growth rate when either population is dilute, set to $1 \text{ day}^{-1}$ to approximate the growth rates of \textit{Synechococcus} \cite{worden2004assessing} unless otherwise noted. The parameter $s$ is the selective advantage of population $A$---population $A$ has a selective advantage $s$ over population $B$ due to differences in competition under crowded conditions when $s>0$. 

To perform a simulation, we first set an initial spatial distribution of populations $A$ and $B$ such that $c_A+c_B=1$ (i.e., the domain is initialized at its no-flow carrying capacity). We then evolve Eqs. \ref{ca} and \ref{cb} forward in time in the presence of the flow field, and measure changes in the distribution and proportion of $c_A$ and $c_B$ after a 24-hour period. Since we are working in the weak compressibility regime (see SI Sec. E for further discussion of this point), the total concentration will remain close to the carrying capacity value as time proceeds ($c_A+c_B \approx 1$).

\subsection{Quantifying community change}
We next provide definitions and conceptual discussion of the two metrics we use to track local and regional changes in community composition.

\subsubsection{Change in relative abundance}\label{dispgrowth}
State of the art high-throughput sequencing technologies quantify microbial community composition using relative abundance \cite{widder2016challenges,sunagawa2020tara,vezzulli2022continuous}. 

The relative abundance of population $A$ in a total population composed of both $A$ and $B$ is defined locally at every point $x$ as
\begin{equation}
f(\*x,t)=\frac{c_A(\*x,t)}{c_A(\*x,t)+c_B(\*x,t)} .\label{deff}
\end{equation}
Normalizing by the initial relative abundance, $f_0$, the change in the spatially-averaged fraction after time $\tau$ is defined as
\begin{equation}
\frac{\Delta \langle f\rangle}{f_0}= \frac{\langle f(t=\tau)\rangle-\langle f(t=0)\rangle }{\langle f(t=0)\rangle},
\label{relabundance}
\end{equation}
where brackets denote spatial averages. 

The change in the spatially-averaged relative abundance measures whether population $A$ becomes more widespread relative to population $B$ after a time $\tau$. Tracking changes in the relative abundance includes the effects of both dispersal and growth/competition, and provides a local measurement of diversity related to $\alpha$-diversity (the number of distinct populations within a local habitat) \cite{whittaker1972evolution, levy2018role}.

\subsubsection{Change in global fraction}
We can gain more information about a population and its regional-scale influence if we also track changes in its biomass. Population $A$ is successful on average over the whole region if its size (i.e. the number of $A$ organisms) increases relative to that of population $B$. 
We define $F^{\text{avg}}$ as the fraction of the total biomass in population $A$ over the whole domain, which we call the global fraction.
\begin{equation}
F^{\text{avg}}(t)= \frac{\langle c_A(\*x,t) \rangle}{\langle c_A(\*x,t)  + c_B(\*x,t)  \rangle},
\end{equation}
where brackets denote spatial averages. Normalizing by the initial value, the change in this global fraction after a time $\tau$ is defined
\begin{equation}
\frac{\Delta F^{\text{avg}}}{F^{\text{avg}}_0}= \frac{F^{\text{avg}}(t=\tau) - F^{\text{avg}}(t=0)}{F^{\text{avg}}(t=0)}.
\label{relgrowth}
\end{equation}

The change in the global fraction can only be nonzero when the growth rate $\mu$ is nonzero, and is unaffected by mixing within the domain. Therefore, tracking the global fraction allows us to evaluate if divergent flows affect the competition between populations and permit differential growth. The global fraction is a global measurement of diversity related to $\gamma$-diversity (the diversity in a broader region). Resolving the absolute abundance is necessary to calculate the global fraction. We note that this measurement is less commonly possible in microbial ecology. 

\begin{figure*} [htp]
    \centering
    \includegraphics[width=0.9\textwidth]{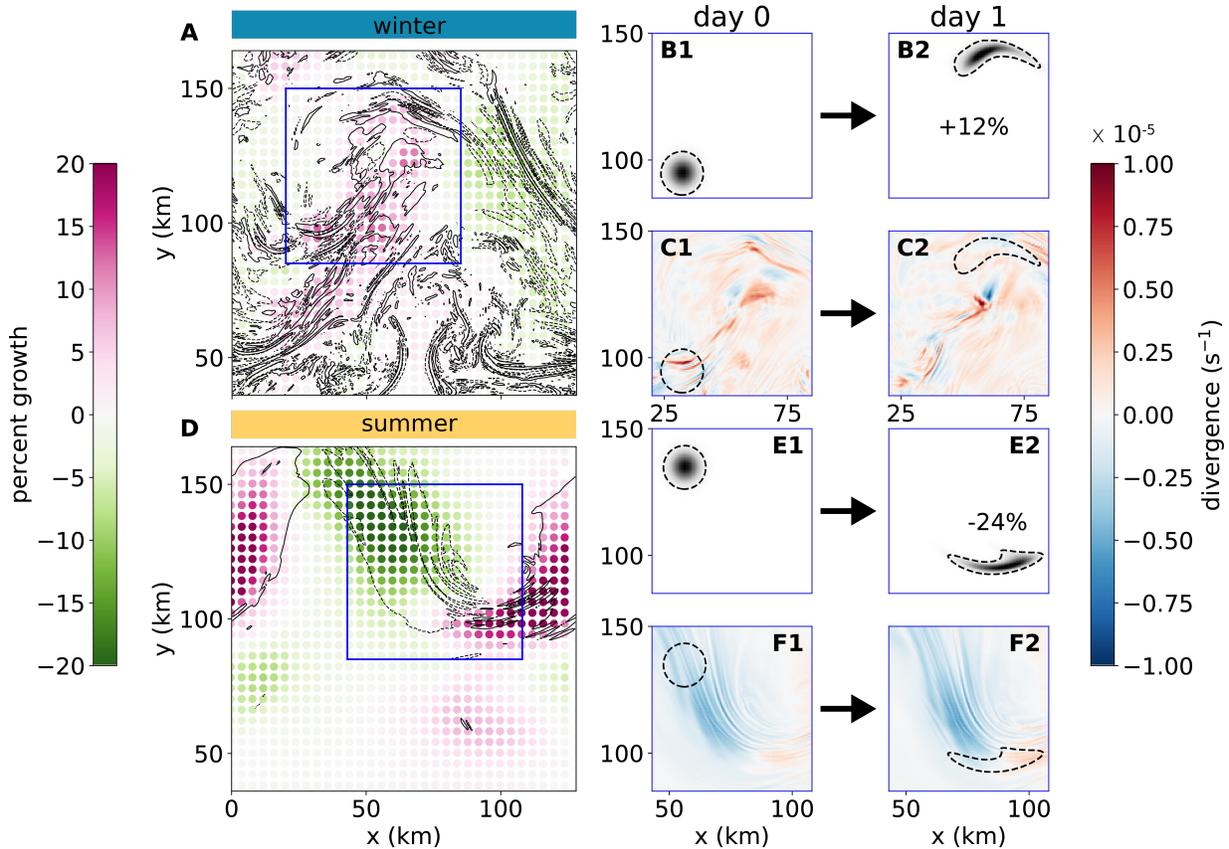}
    \caption{The spatial distribution of velocity divergence affects the growth (measured here as the change in the relative abundance) of local populations, shown in Figures A and D by the qualitative agreement between divergence contours and regions of positive/negative growth. Panels A--C use the winter flow field. Panels D--F use the summer flow field. (A,D) The change in the relative abundance, expressed as a percent (100 times Eq. \ref{relabundance}) between the final and initial populations of type $A$ after one day. Each dot is located at the spatial center of the localized population’s Gaussian initial condition, with the color giving the magnitude of the change. The contours show divergence equal to $10^{-6} \text{s}^{-1}$ (solid) and $-10^{-6} \text{s}^{-1}$ (dashed). The blue boxes show the spatial extent of the subdomains plotted in panels B,C and E,F. (B,E) Concentration of a example localized populations of type $A$ as a function of space at the initial and final time. Black is concentration equal to one, and the black dashed lines contour where the population concentration $c_A$ is equal to 0.1. (C,F) Divergence as a function of position in the velocity field at the initial and final time, with the same black dashed lines as in panels B and E. In these trials, a diffusivity of $5 \text{ m}^2\text{s}^{-1}$ is used.}
    \label{maps}
\end{figure*}

\section{Results}

Motivated by our observations of variations in the relative abundance of oligotypes which we assume to be neutral competitors, as well as our desire to isolate the effects of flow from the effects of selection, we first set the selective advantage $s=0$. 
We initialize the system such that population $A$ is localized according to a Gaussian distribution centered on a particular $x,y$ coordinate with a standard deviation of 4 km. We set the concentration of population $B$ such that $c_B=1-c_A$ everywhere. Therefore, the total concentration is everywhere equal to the equilibrium carrying capacity in the absence of flow. Each dot in Fig. \ref{maps} A,D represents an independent trial initialized in this manner, with the Gaussian population centered at a different location. 

This numerical experiment reveals the main theoretical and computational results of this report.
After evolving a localized population of a neutral competitor for 24 hours, we find that the community composition can be significantly affected by regions of divergence in the flow field. Regions of positive divergence disperse organisms, locally decreasing competition and stimulating growth. Notably, the population does not pass through regions of positive and negative divergence quickly enough that the accrued advantages/disadvantages average to zero.

The divergent flows affect both the local community composition --- quantified as the relative abundance (Eq. \ref{relabundance}) --- and regional community composition --- quantified as the global fraction (Eq. \ref{relgrowth}). The regions of positive and negative divergence lead to $\sim \pm 20\%$ changes in the relative abundance of population $A$ without appreciably changing the total biomass (SI Sec. E).  Since non-divergent flows cannot alter spatially averaged relative abundances or the global fraction in a closed system at its carrying capacity in the absence of noise and selection (SI Sec. F), these changes can be attributed to the effect of the regions of divergence.



We next explore these results, displayed in Fig. \ref{maps} and \ref{c_correlation}, in more detail and compare with theoretical expectations.

\subsection{Local influence of divergence}

\begin{figure} [htp]
    \centering
    \includegraphics[width=\columnwidth]{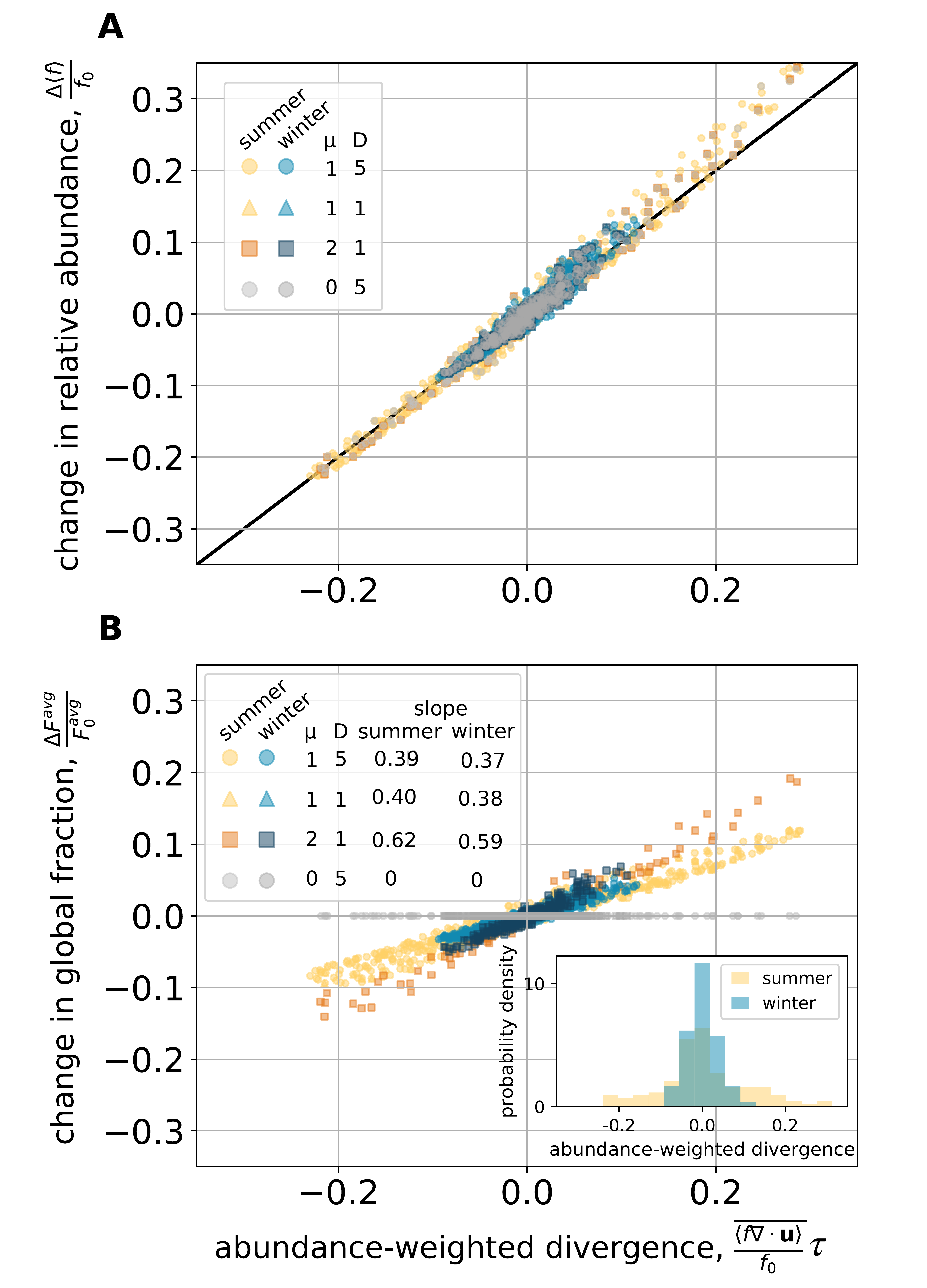}
    \caption{Changes in the relative abundance and global fraction of a localized population are strongly dependent on the local flow conditions. For a given model and parameter combination, each point represents a different initial population location. 
    (A) The normalized change in the relative abundance of population $A$ (Eq. \ref{relabundance}) over one day as a function of the integrated divergence experienced by that population. The solid black line shows the 1:1 line (our theoretical expectation, Eq. \ref{s0fracchange}). (B) The change in the global fraction of population $A$ (Eq. \ref{relgrowth}) over one day as a function of the integrated divergence experienced by that population. Inset shows the distribution of population-weighted integrated divergence experienced by populations initialized across the domain in the two different models. 
    The $\mu \neq 0$ populations evolved in the summer and winter flow fields are shown in yellow/orange and blue/indigo, respectively. The symbol shape denotes different trial parameters---diffusivity D (m$^2$s$^{-1}$) and population growth rate $\mu$ (day$^{-1}$) are varied. The $\mu=0$ simulations are given grey markers. In (B), we see that $\Delta F^{\text{avg}}=0$ for the $\mu = 0$ trials, as expected. The integrated abundance-weighted divergence is computed using snapshots of the population and flow field taken every three hours. Between 144 and 1056 points for each model and parameter combination are shown.}
    \label{c_correlation}
\end{figure}

We observe that changes in the spatially-averaged relative abundance have a linear relationship with the divergence experienced by that population, averaged over space and time in Fig. \ref{c_correlation}A. 

The linear relationship between relative abundance and divergence holds even when the organisms are not able to reproduce ($\mu=0$). In this case, changes in the relative abundance are solely due to dispersal. For an intuitive example of how Eq. \ref{relabundance} can be nonzero in the absence of growth, consider a region of positive divergence that is occupied solely by a non-reproducing population $A$. The flow will distribute $A$ organisms throughout the system, increasing the relative abundance of $A$ outside the source region. At the source itself, the relative abundance will remain locally equal to 1 as long as no $B$ organisms are introduced, despite the local depletion in the amount of $A$ organisms. Thus, the spatially averaged relative abundance can increase or decrease even when there is no growth. When $\mu \neq 0$, the local depletion will be compensated by growth at the source, maintaining an approximately uniform distribution of biomass. 

Since the $\mu=0$ trials follow the same trend as the $\mu \neq 0$ trials, we conclude that the observed increase in relative abundance over this 24-hour time period is primarily due to dispersal, rather than differential growth.

\subsubsection{Relative abundance in the weak compressibility limit}\label{theory}
We can calculate the relationship between velocity divergence and changes in the relative abundance in the weak compressibility limit. 
From the equations for the evolution of the concentrations of populations $A$ and $B$ (Eqs. \ref{ca} and \ref{cb}), we obtain an equation for the evolution of the relative abundance of $A$ (SI Sec. F).
\begin{equation}
    \frac{\partial f}{\partial t} + \*u \cdot \nabla f = D \nabla^2 f + \frac{2 D}{c} \nabla f \cdot \nabla c +  s \mu c f(1 - f),
    \label{eq:frac}
\end{equation}
where $c(\*x,t)=c_A(\*x,t)+c_B(\*x,t)$ is total concentration, expressed as a fraction of the local carrying capacity. In this equation, the growth rate $\mu$ only appears directly in a logistic competition term, and implicitly as the relaxation rate of $c$. The second term on the right hand side of Eq. \ref{eq:frac} is small for the case of a weakly compressible flow (SI Sec. D and \cite{plummer}). Therefore, by integrating by parts and setting $c=1$ in the selection term, the rate of change of the relative abundance integrated over space in a weakly compressible flow can be approximated 
\begin{equation}
     \frac{\partial}{\partial t}\int_\Omega f d\Omega \approx \int_\Omega \left[ f \nabla \cdot \*u + s \mu f(1 - f) \right]d\Omega,
    \label{eq:frac2}
\end{equation}
where $\Omega$ is the area of the 2D domain. Note that the boundary terms can be neglected when population $A$ is localized, as in our simulations (SI Sec. F). 

To compare the behavior of this equation with simulations, we integrate with respect to time and divide both sides by $f_0 \Omega\equiv\int_\Omega f(t=0) d\Omega$. 
\begin{align}
   & \frac{\frac{1}{\Omega}\left(\int_\Omega f(t=\tau) d\Omega\right)- f_0 }{f_0 } \approx \nonumber \\& \int_{0}^{\tau} \left( \frac{1}{\Omega}\int_\Omega \left[\frac{ f}{f_0} \nabla \cdot \*u + s \mu \frac{f}{f_0}(1 - f) \right]d\Omega\right)dt.
    \label{fulleq}
\end{align}
For the case of neutral competition, we set $s=0$, and Eq. \ref{fulleq} reduces to 
\begin{equation}
\frac{\Delta \langle f\rangle}{f_0}\approx \frac{\tau}{f_0} \overline{\left\langle f \nabla \cdot \*u \right \rangle } ,
    \label{s0fracchange}
\end{equation}
where the brackets denote averages over all space, and the overbar denotes an average over the time interval $t=0$ to $t=\tau$. 

\subsubsection{Agreement between simulations and theory}
There is excellent agreement between the theory (Fig. \ref{c_correlation}A, black line) and simulations for different flow fields, growth rates, and diffusivities (a diffusivity greater than the carrier fluid diffusivity could represent, for example, active dispersal). As expected from Eq. \ref{s0fracchange}, there is no obvious dependence on the growth rate $\mu$ or diffusivity $D$. This agreement shows that the weak compressibility assumption that we made to derive Eq. \ref{s0fracchange} is reasonable for these flow fields. 

Over longer time periods we would expect that the results of the experiments with $\mu > 0$ to differ more from those with $\mu = 0$. Concentration gradients, $\nabla c$, will become large for the $\mu=0$ simulations, violating the assumption of weak compressibility. Sufficiently large growth prevents the development of strong gradients in concentration.



We note that we display relatively fewer data points for strong negative divergence, as these trials were most susceptible to numerical instability. 



\subsection{Regional influence of divergence}
We examine the relationship between divergence and changes in the global fraction to disentangle the effects of dispersal from the effects of growth and competition.

We find that changes in the global fraction have a linear dependence on the divergence experienced by that population (Fig. \ref{c_correlation}B), although the slope of the trend is smaller than for changes in the relative abundance.

This discrepancy would not occur if $c_A+c_B$ were strictly equal to 1 everywhere, in which case the global fraction and spatially-averaged relative abundance would be identical.  However, even when $c_A+c_B \approx 1$, as is the case for the weakly compressible flows considered here, there can be significant differences between these two measures. Even if on average the domain is uniformly occupied, especially strong convergences and divergences cause small local accumulations and deficits, which must be taken into account to understand the change in the global fraction, $\Delta F^{\text{avg}}/F^{\text{avg}}_0$, and the influence of competition and growth on changes in population distributions. As a result of these concentration fluctuations, the global fraction has a weaker dependence on the divergence than does the relative abundance. 

\subsubsection{Fluctuations about the weak compressibility limit}
To understand how the global fraction can differ from the spatially-averaged relative abundance, we consider the equation for the change in the total concentration (SI Sec. F). 
\begin{equation}
     \frac{\partial c}{\partial t} + \nabla \cdot  (\*u c) = D \nabla^2 c + \mu c (1- c).
\end{equation}
We model a small fluctuation in the total concentration, setting $c=1+\epsilon$ and assume that the growth is much larger than the divergence ($\mu \gg \nabla \cdot \*u$). We neglect the time derivative as in \cite{perlekar2010population}, and drop terms proportional to $\nabla \epsilon$ and $\epsilon (\nabla \cdot \*u)$ to obtain
\begin{equation}
    \epsilon\approx- \frac{1}{\mu} (\nabla \cdot \*u).
\end{equation}
With this approximation, the spatially integrated relative abundance becomes
\begin{equation}
  \int_\Omega f d\Omega \approx \int_\Omega \frac{c_A}{1- \frac{1}{\mu} (\nabla \cdot \*u)} d\Omega \approx \int_\Omega \left( c_A + \frac{c_A (\nabla \cdot \*u)}{\mu} \right) d\Omega.
\end{equation}
Since we expect $c_A$ to depend on $\nabla \cdot \*u$ when $c$ is allowed the fluctuate, the second term will not integrate to zero.

Upon substituting this expression in to Eq. \ref{eq:frac2}, integrating with respect to time, noting $F^{\text{avg}}(t) \approx \int_\Omega c_A d\Omega/\Omega$, and taking the case of no selective advantage for simplicity, we find
\begin{equation}
    \frac{\Delta F^{\text{avg}}}{F_0^{\text{avg}}} \approx \frac{\tau}{F_0^{\text{avg}}}\overline{\langle f \nabla \cdot \*u\rangle} - \frac{\Delta \langle c_A \nabla \cdot \*u \rangle}{\mu F_0^{\text{avg}}},
    \label{favg_eqn}
\end{equation}
where $\Delta$ indicates a difference between the initial and final time points. By modeling a fluctuation in the total concentration, we thus observe that the change in the global fraction, unlike the change in relative abundance, has an explicit dependence on $\mu$ that goes to zero when $\mu \to \infty$, at which point $c$ is strictly equal to $1$. Due to the approximations made, this relation breaks down for small $\mu$.

\subsubsection{Agreement between simulations and theory}
We observe a $\mu$ dependence in the relationship between the global fraction and the weighted divergence, as expected from the fluctuation model of Eq. \ref{favg_eqn}, but no clear dependence on the flow field (winter vs. summer) or diffusivity (Fig. \ref{c_correlation}B). There is no change in the global fraction when there is no growth ($\mu=0$), as in that case all covariance between divergence and $c_A$ is due to accumulation. Higher values of $\mu$ produce trends closer to the one-to-one relationship of Fig. \ref{c_correlation}A, as expected. The slopes of the lines of best fit for each set of parameters are given in the figure legend. Due to the approximations made in the fluctuation model, Eq. \ref{favg_eqn} cannot be used to quantitatively predict these slopes.

\subsection{Quantifying the effect of divergence via selective advantage}\label{sec:intruder}

Using the case of neutral competition in the previous section allowed us to isolate the effects of the particular physical-biological mechanism revealed in this study. However, the influence of divergence on community composition can be large enough that it can allow a population to overcome a selective disadvantage. By studying the interplay between divergence and selective advantage, we also are able to quantify the effect of a region of divergence on a population in terms of an effective selective advantage.

We consider a simulation in which a localized population is initialized in a region of strong positive divergence, selected using the data from the trials in Fig. \ref{c_correlation}. This population was found to experience growth and increased abundance relative to its competitor (i.e. it corresponds to a point in the top right corner of Fig. \ref{c_correlation}A,B). We now alter the competition dynamics by imposing a selective advantage/disadvantage on the population (nonzero $s$ in Eqs. \ref{ca} and \ref{cb}). For some negative value of $s$, $s^*$, the selective disadvantage will exactly counterbalance the effect of the positive divergence, and the spatially averaged relative abundances will not change over our observation period. For $s< s^*$, the relative abundance of species $A$ will decrease. We therefore consider $|s^*|$ to be the {\em effective selective advantage} provided to the population by the flow field. 

This simulation can be thought of as tracking, for example, a low-light specialist organism arriving at the surface via an upwelling, where it is comparatively ill-suited to survive. A sufficiently strong upwelling underneath a disadvantaged ecotype could act as a lifeline, and allow it to avoid competitive exclusion in its newly harsh environment. Note that at the surface, divergent flow generates upwelling in the vertical, but this exact correspondence between strong upwelling (vertical velocity) and strong 2D divergence (vertical velocity gradient) does not necessarily hold subsurface. 

\begin{figure}[htp]
    \centering
    \includegraphics[width=\columnwidth]{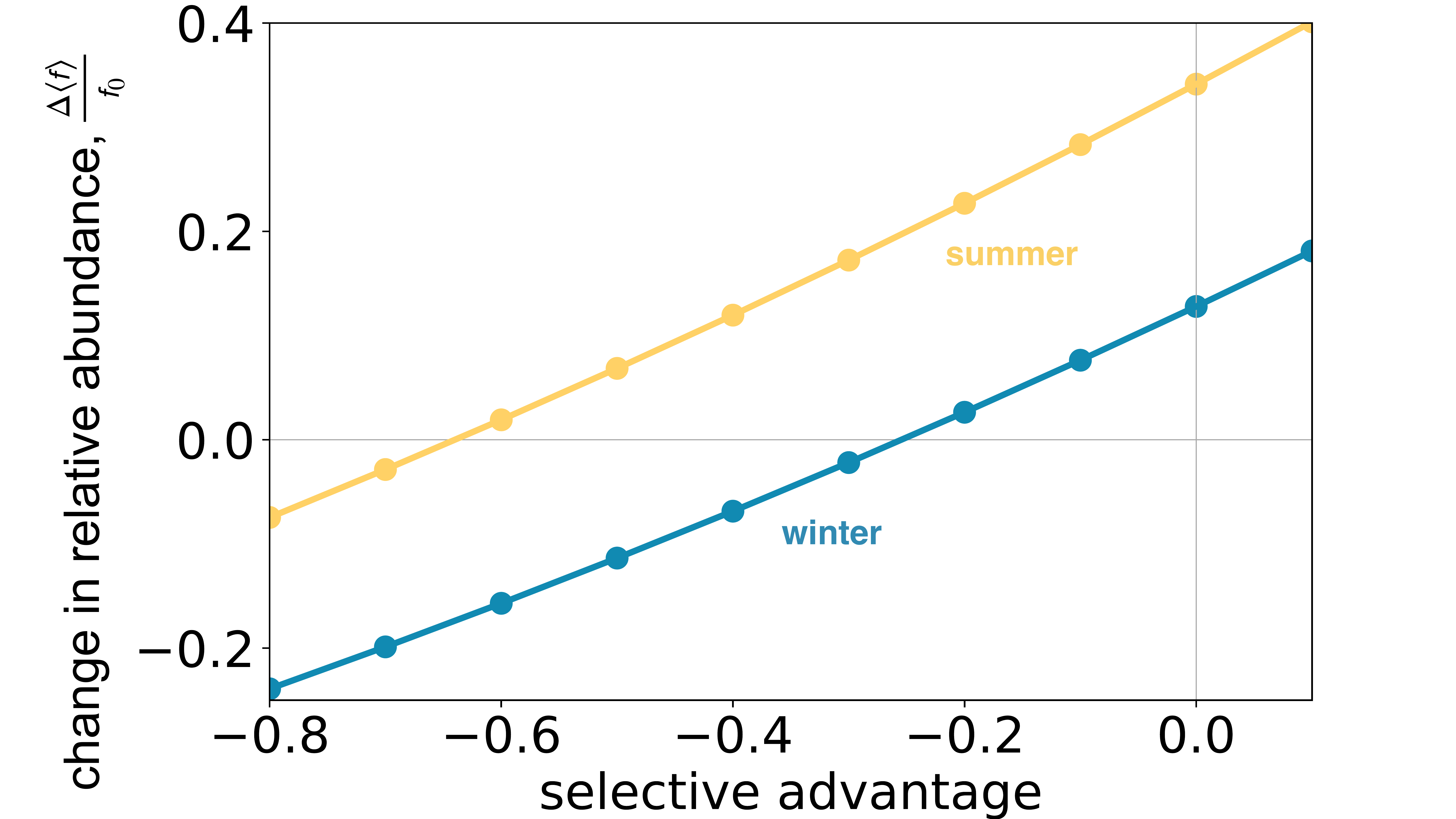}
    \caption{Change in relative abundance over 24 hours as a function of the selective (dis)advantage. The blue data correspond to the population shown in Fig. \ref{maps}B: the initially Gaussian population that yielded the greatest value of $\Delta \langle f\rangle/f_0$ in the neutral simulations of Fig. \ref{c_correlation}A for the winter flow field. The yellow data show results for the corresponding initialization for the summer flow field. A selective disadvantage of approximately 0.65 for the summer flow field and 0.25 for the winter flow field is required to cancel out the effect of the divergence experienced by the population.}
    \label{fig:fig4}
\end{figure}

In Fig. \ref{fig:fig4}, we observe that the winter flow field can compensate for a selective disadvantage of $s^*=-0.25$, and the summer flow field can compensate for a selective disadvantage of $s^*=-0.65$. Since we place the localized populations at sites that we know are particularly advantageous in these simulations, $s^*$ should be thought of as a maximum effect size under idealized conditions. In SI Sec. G, we present a simplified theory that provides a reasonable estimate for both the slope and x-intercept, $s^*$, of the trends in Fig. \ref{fig:fig4}.

\subsection{Realistic population structures}\label{sec:selection}

In the ocean, populations are rarely spatially localized and instead display correlations with water masses and dynamics, due to physical, chemical, and ecological factors. To study the implications of divergence on spatially extended plankton communities, we designed three initial conditions based on realistic phytoplankton biogeography for the summer flow field. These initial conditions exemplify a few ways that ecological communities may be distributed relative to a front, where the effects of compressibility are the largest. 

\begin{figure}
    \centering
    \includegraphics[width=\columnwidth]{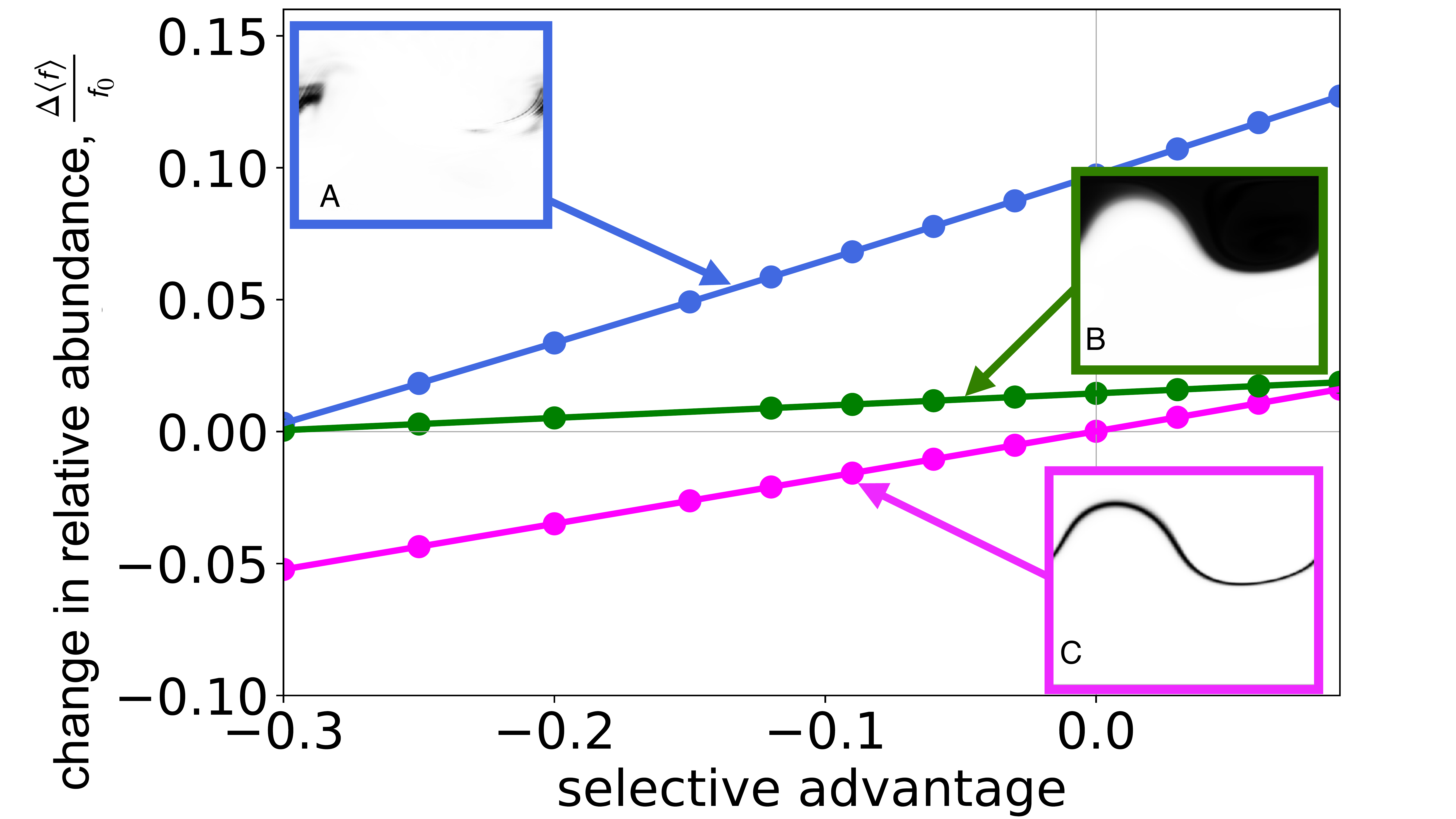}
    \caption{Change in relative abundance of the spatially extended populations over 12 hours of the summer flow field as a function of the selective (dis)advantage, as in Fig. \ref{fig:fig4}. The initial concentration profiles of population $A$ are pictured in the insets with $A$ representating the whole concentration in areas shaded black and none in areas shaded white. Inset (A) shows the  upwelling community (slope =  0.32), (B) shows distinct populations in distinct water masses (slope = 0.05), and (C) shows a frontal community (slope = 0.18).}
    \label{fig:fig5}
\end{figure}

First, we might expect an upwelling to carry new species to a given depth level, as discussed in the previous section \cite{stanley2017submesoscale}. The upwelling-inspired initial condition (shown in Fig. \ref{fig:fig5}A) is constructed by placing population $A$ in all areas with upwelling, while still requiring $c_A+c_B=1$. We observe that the distribution of regions of positive divergence is much more complex than a simple Gaussian initial condition. Second, the water masses that meet at a front will often have distinct communities \cite{d2010fluid,clayton2013dispersal}. The distinct water mass initial condition, shown in Fig. \ref{fig:fig5}B, is defined by placing population $A$ in the region where the salinity is higher than 36.5 PSU and population $B$ in the rest of the domain. Third, another possibility is that the front has a unique community due to influences of frontal currents on the rate of nutrient supply \cite{nagai2019multiscale,palter2020high}. This frontal initial condition is constructed by placing population $A$ near 36.5 PSU, as shown in Fig. \ref{fig:fig5}C. We evolve these initial conditions in the summer flow field, with the selective advantage/disadvantage of the community varied to measure the effective advantage conferred by the flow. These simulations were only evolved for 12 hours because the fine structure in the initial conditions made them more susceptible to numerical instabilities. We expect that doubling the period of the simulation would double the slope of the trends in Fig. \ref{fig:fig5}, while keeping $s^*$ constant. 

The distribution of the population affects the change in the population relative abundance. When there is no selective advantage ($s = 0$), the population initialized in an upwelling experiences the greatest change in relative abundance (Fig. \ref{fig:fig5}A; 9.7\% change over 12 hours). The flow field is near the surface so there is a nearly linear relationship between upwelling and divergence. The two populations that are defined by salinity criteria, the frontal population (Fig. \ref{fig:fig5}C) and the population on the dense side of the front (Fig. \ref{fig:fig5}B) have smaller changes in relative abundance when the two populations are neutral competitors with a change of 0.01\% over 12 hours for the frontal population and 1.5\% over 12 hours for the population on the dense side of the front.
 However, the population initialized on the dense side of the front has a similar effective selective advantage ($s^* \approx -0.32$) to the population initialized in the upwelling region ($s^* \approx -0.3$). Although the disperal advantage imparted by the flow to the neutral population is relatively small for the population on the dense side of the front, the spatial population structure allows this population to overcome a large disadvantage.

We can gain insight into the relationship between selective advantage and change in relative abundance by including selection in Eq. \ref{s0fracchange}:
\begin{equation}
\frac{\Delta\langle f\rangle}{f_0} \approx \frac{\tau}{f_0}(\overline{\langle f \nabla \cdot \mathbf{u}\rangle}+s \mu \overline{\langle f(1-f)\rangle}).
\end{equation}
We therefore expect the change in relative abundance as a function of selective advantage to depend on the spatial distributions of the populations. The regions where $f(1-f)$ is nonzero are where the two populations are in contact with each other and therefore competition is most important. The size of these regions sets the slope of $\Delta \langle f \rangle/f_0$ versus $s$. The selective disadvantage at which the relative abundance does not change, $s^*$, depends on both the spatial distributions of the populations and the divergence of the flow. As shown in Fig. \ref{fig:fig4}, two populations with the same initial condition in different flow fields have the same dependence on the selective advantage (the slopes), but different values of $s^*$ (the x-intercepts). In Fig. \ref{fig:fig5}, we see that populations in the same flow field with different initial conditions have different slopes and different x-intercepts.

\section{Discussion}
Motivated and informed by observations, our coupled biophysical model considers realistic oceanic flow fields, resolved at the submesoscale, acting on phytoplankton populations restricted to live within a particular depth range. Phytoplankton living at regions of positive divergence enjoy the advantage of having would-be competitors constantly swept away by the flow, allowing offspring to easily spread. Those living at regions of negative divergence are instead challenged by a stream of new arrivals.

The realistic oceanic flow fields are in a regime where the effect of divergence is primarily dispersal rather than loss of biomass \cite{perlekar2013cumulative, plummer}, resulting in variations in relative abundance of up to 35\% over one generation. Nonetheless, the effects of compressibility also affect growth and competition, resulting in variations in the global fraction of up to 20\%. Here we have shown that these simulated trends are consistent with theoretical expectations and of the same magnitude as the observed variations in community structure at a front in the Mediterranean Sea.  The effects of divergence are integrated over time and are therefore stronger when a population resides in an area of positive divergence for a longer time. In these simulations, the summer flow field has a simpler divergence structure which leads to larger divergence when integrated over a day.

The results suggest a mechanism that can affect plankton biogeography, alongside other established mechanisms such as fluctuations in light, temperature, and ecological interactions \cite{richerson1970contemporaneous}. The impact of effective compressibility may vary on long space and time scales because velocity divergence patterns display seasonality and regional variation \cite{callies2015seasonality,choi2017submesoscale}. Our calculations and simulations assume a uniform nutrient distribution and consequently constant carrying capacity---if nutrients had been modeled explicitly, the advantage afforded by regions of positive divergence could be enhanced at the surface, with organisms born in these regions experiencing an even greater advantage due to the associated upwelling supplying nutrients \cite{perruche2011effects,freilich2021diversity}. We note that all of the effects discussed in this work would also arise in the more general oceanographic case where organisms are restricted to remain close to a fixed density surface rather than a fixed depth (realistic for organisms that may regulate their buoyancy, for example \cite{guasto2012fluid}). This scenario is discussed in SI Sec. D.

In some oceanographic observations, regions of velocity convergence have been shown to result in accumulation of phytoplankton populations. For example, convergence has been observed to impact biological populations in a coastal region of the Western Mediterranean Sea \cite{hernandez2018effect} and in open ocean regions \cite{guidi2012does,palter2020high,benavides2021fine}. These convergence zones may hold particular biogeochemical significance because they disproportionately cause accumulation of buoyant phytoplankton such as nitrogen-fixers \cite{palter2020high,benavides2021fine}. 

The numerical experiments highlight that effective compressibility, unlike many mechanisms by which advection can affect competition, can be relevant even when populations are ecologically neutral (equally matched competitors) in the absence of a flow. Neutral theories in ecology emphasize the role of stochasticity and dispersal on population dynamics. These dynamics can be consequential because even if plankton types are neutral in their competition under the conditions at a given moment in time, they may differ in other ways, which means that the outcomes of the neutral competitions have biogeochemical implications \cite{treguer2018influence}. 

Incorporating divergence of the flow field into analysis of ecological processes may have relevance beyond the processes studied here. For example, positively buoyant artificial particles such as microplastics may accumulate in regions of convergence either at the mesoscale \cite{baudena2022streaming} or at the gyre scale \cite{law2010plastic}. These microplastics have microbial communities associated with them \cite{amaral2020ecology}. However, the framework of weak compressibility only applies when population growth is sufficiently fast to maintain an approximately uniform total concentration. Deriving results for a strong compressibility regime and applying them to populations of buoyant artificial particles may be a fruitful avenue for future research.

Effective compressibility can either promote or suppress diversity, depending on the population structure. In Sec. \ref{sec:intruder}, for example, we demonstrate that a positive divergence can compensate for a competitive disadvantage. If rare species often occupy regions of positive divergence (for example, if they are brought to the surface by an upwelling event), effective compressibility should increase diversity. If instead rare plankton populations are drawn to downwellings, where their populations are more likely to shrink, diversity will be suppressed. 

The proposed mechanism and results presented here cannot be quantitatively validated using the existing observations. However, the sign and magnitude of the change in the community composition between the two observed transects is close in magnitude to the convergence observed in the transect in the Western Mediterranean Sea, supporting the idea that divergence could alter the community composition. The overall convergence is approximately $2\times 10^{-5}$~s$^{-1}$ and the decrease in the relative abundance of the population from the upstream to the downstream transect (advection time of $\sim 3$~hours between the transects) is 5--10\%. 

\section{Conclusions}
This work introduces a novel mechanism for the specific influence of submesoscale flows on plankton communities. The combined observational, computational, and theoretical evidence suggests that horizontal velocity divergence can generate measurable variations in community composition through modulation of ecological competition.  Thus, submesoscale dynamics can profoundly influence local and regional changes in community composition and biodiversity of the subtropical communities responsible for the nitrogen cycle, carbon dioxide uptake, and primary production. 

At the scale of ocean fronts, phytoplankton can experience weakly divergent flows that disperse plankton populations and alter competition and growth. Both simulations and theory support the conclusion that regions of divergence significantly affect both the spatially-averaged relative abundance as well as the global fraction (when the growth rate is nonzero), though the details of the relationships differ. Regions of positive divergence support local populations, while regions of negative divergence suppress them. The regions of divergence in the flow fields examined here can lead to differences in relative abundances of up to 35\% in ecologically neutral populations over a 24-hour period, which is consistent with the magnitude of population change in observations. These divergent flows provide an effective selective advantage of up to 65\%. The effect of divergence is most likely to be a dominant driver of demographic change in locations with strong divergence, which occur over timescales of hours to days, and when oceanographic or physiological factors confine organisms to a given depth range. Divergence (i.e. effective compressibility) should be considered as a potential additional explanation for patchiness in community composition. 

This study is limited to considering the effects of divergence alone without other variations in the growth environment, such as nutrient supply. The joint effects of submesoscale variations in growth and loss processes and this mechanism of divergence could be a fruitful area for future research. More rigorous observational confirmation requires Lagrangian observations of microbial diversity, which will likely be technologically feasible in the near term. Observations should quantify divergence, population growth and selective advantage, and ecological effects. The processes discussed here are likely most important for positively buoyant populations and populations that are depth-stratified through other mechanisms in frontal regions where divergence is relatively large. 

\section*{Acknowledgments}
The authors thank David R. Nelson and John Toner for useful discussions, Camille Poirier and Sebastian Sudek for assistance with biological sample processing, Eva Alou and Andrea Cabornero for providing the nutrient samples, John Allen for processing the VM-ADCP observations, Eric D'Asaro for serving as co-chief scientist of the research cruise, Mathieu Dever, Sebastian Essink, Kausalya Mahadevan, and Alex Beyer for sampling assistance at sea, and the captain and crew of the NRV Alliance for their assistance and expertise. Funding was provided by a Montrym grant and Martin Fellowship from MIT.

\section*{Author Contributions}
A.P., M.F., R.B., F.T., and A.M. designed research; A.P., M.F., C.J.C., and L.S. performed research; A.W. contributed new reagants or analytic tools; A.P. and M.F. analyzed data; A.P. and M.F. wrote the paper; A.P., M.F., R.B., A.W., F.T., and A.M. revised the paper.

\section{Open Research}
The model output and biological model code is available on Zenodo \cite{datarepo}. The code to generate the oceanography flow fields is available on Zenodo \cite{psomrepo} (PSOM v1.0 with initial conditions in the released code).
The sequences are available with BioSample accession numbers SAMN28021319-SAMN28021334. This study has been conducted using E.U. Copernicus Marine Service Information \cite{aviso}.

%

\end{document}


\title{Supporting Information S1}

\maketitle
\vspace{-3em}
\tableofcontents{}
\vspace{-1em}

\subsection{Oceanographic flow fields}
The edges of the Alborán Gyres (which include the Almería-Oran front) are persistent density fronts between the Mediterranean and Atlantic water masses in the Western Mediterranean. The front is modified by instabilities at the mesoscale and submesoscale, which generate regions of divergence and convergence. There is a strong along-front flow as well as an ageostrophic cross-front flow \cite{mahadevan2016impact}.
\subsubsection*{Model}
\begin{figure}
    \centering
    \includegraphics[width = 0.55\textwidth]{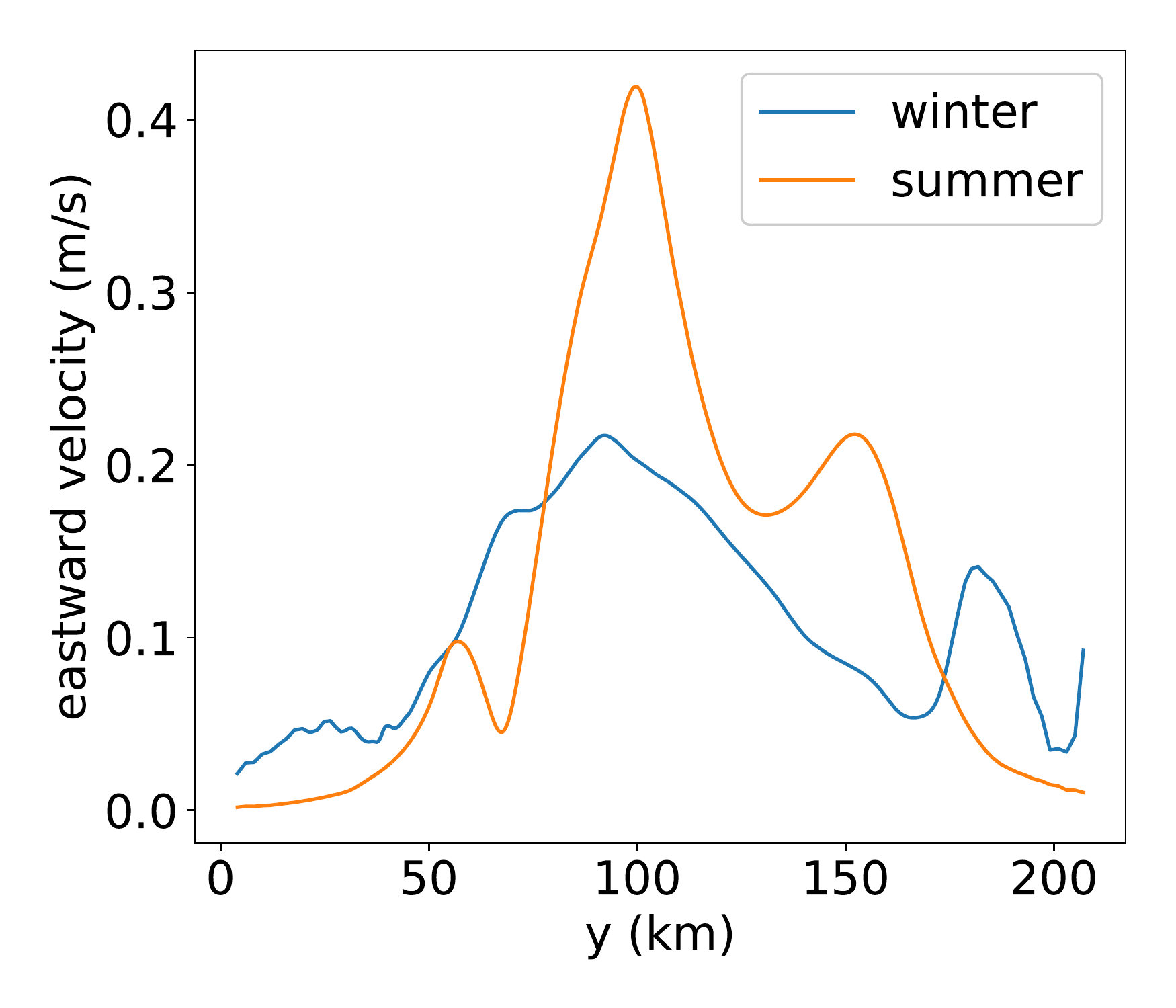}
    \caption{Root mean-square jet speed in the periodic (east-west) direction, averaged over the 24-hour period used in simulations. Positive values are eastward.}
    \label{fig:u_vel}
\end{figure}

The flow fields are dominated by an eastward flowing jet that is more variable in the winter model than in the summer model. The mean jet speed is therefore roughly twice as fast in the summer as in the winter (Fig. \ref{fig:u_vel}). Over a 24-hour period, the frontal jet present in the winter flow field moves a localized population at most a horizontal distance corresponding to approximately 14\% of the domain, and the summer flow field moves the population at most a distance of approximately 28\% of the domain. 

\begin{figure}
    \centering
    \includegraphics[width = \textwidth]{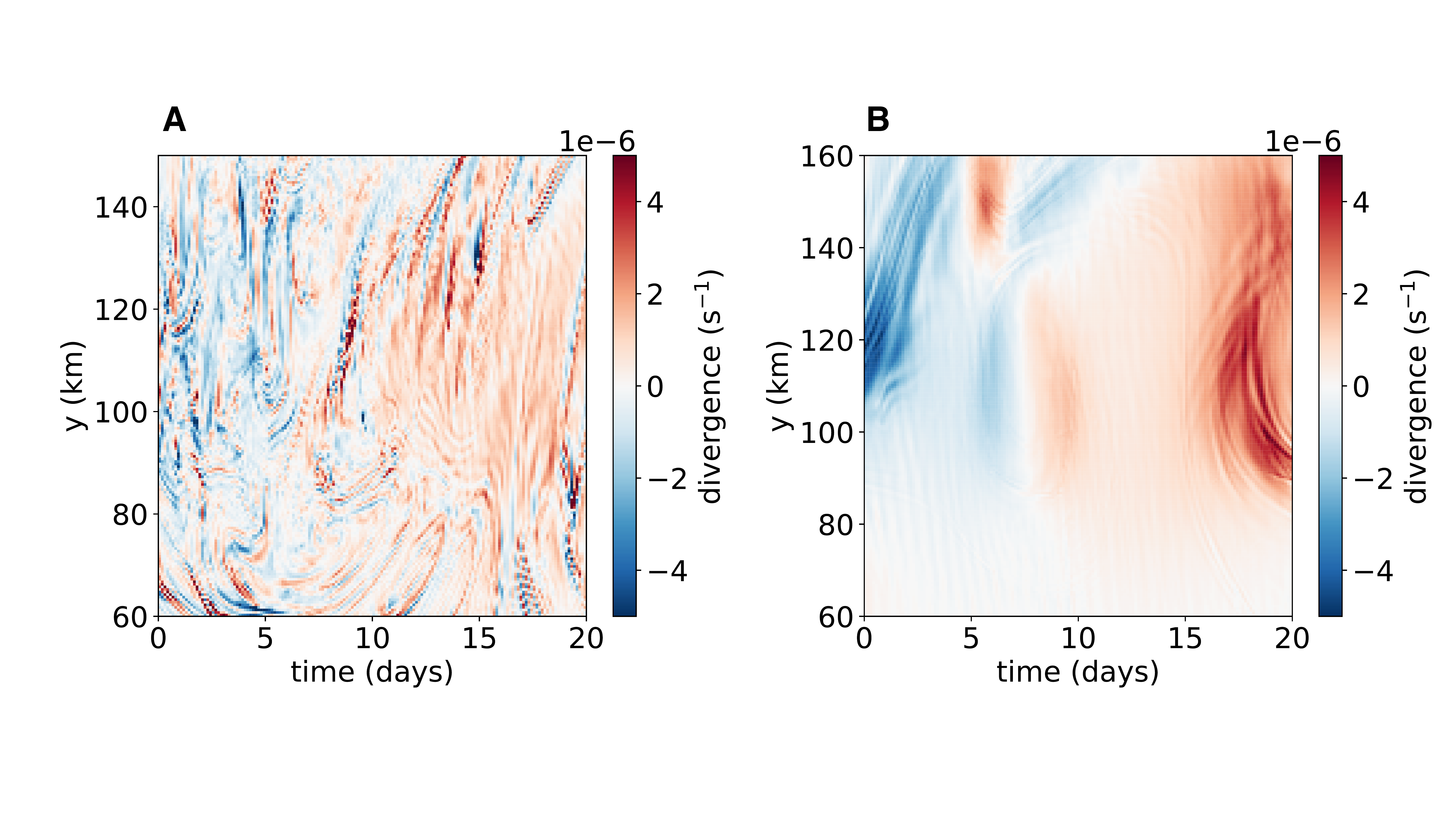}
    \caption{Hovm\"{o}ller diagrams of the divergence in (A) the winter flow field and (B) the summer flow field.}
    \label{fig:hov_div}
\end{figure}

The large scale structure of the jet dictates the locations of divergent regions in the summer model (Fig. 2F). The regions of divergence in the winter model are also influenced by the meandering jet location, but have finer scale structure as well (Figs. 2C and \ref{fig:hov_div}).

\begin{figure}
    \centering
        \includegraphics[width = 0.55\textwidth]{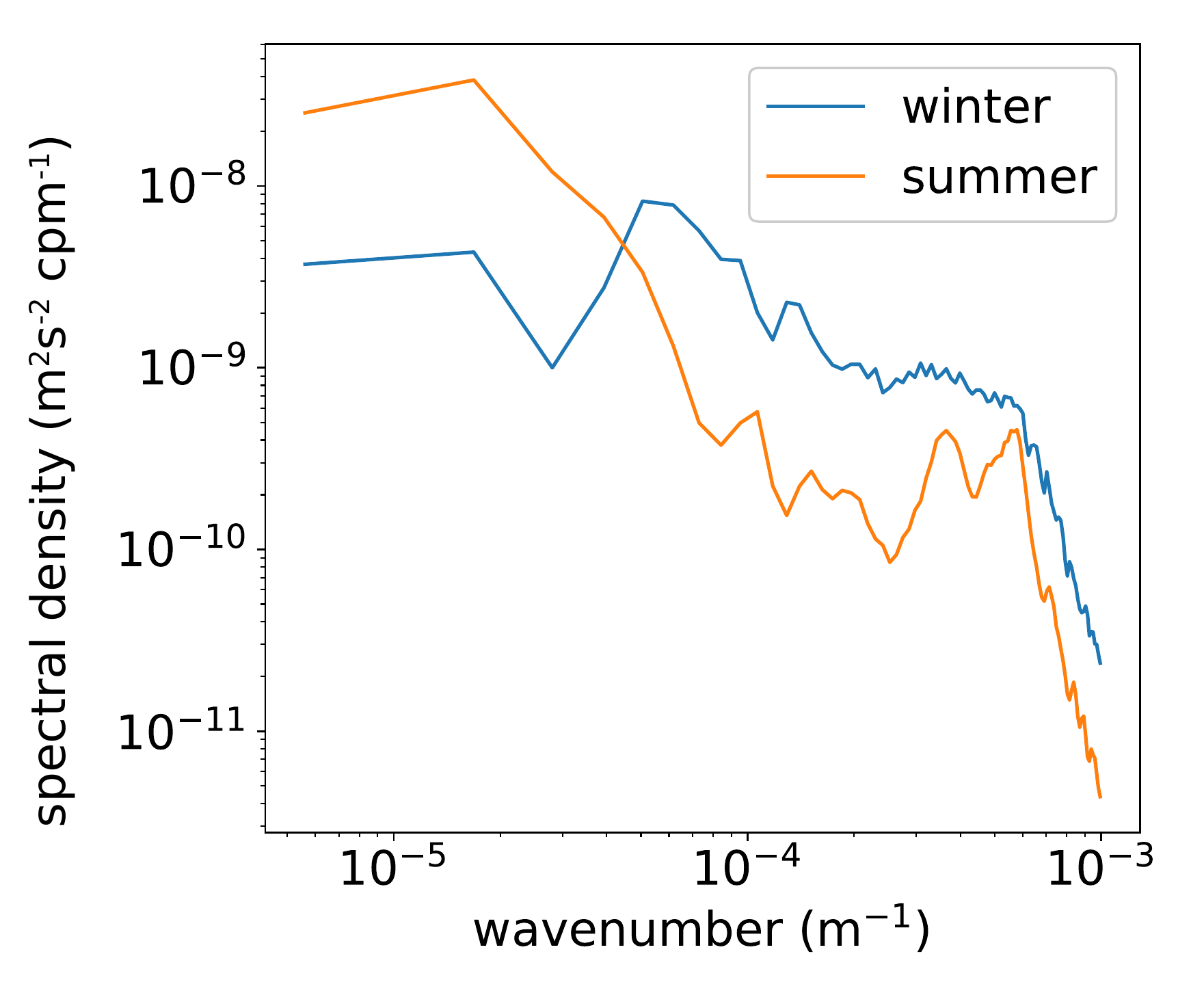}
   \caption{Wavenumber spectrum of the divergence during the simulation day from both models.}
    \label{fig:spec_div}
\end{figure}

Given this spatial structure, the divergence has higher power spectral density in the summer model than the winter model at large spatial scales (low wavenumber) but falls off more rapidly such that the divergence has higher power spectral density at intermediate and small spatial scales in the winter model (Fig. \ref{fig:spec_div}). 

\subsubsection*{Observations}
\begin{figure}
    \centering
    \includegraphics[width = 0.8\textwidth]{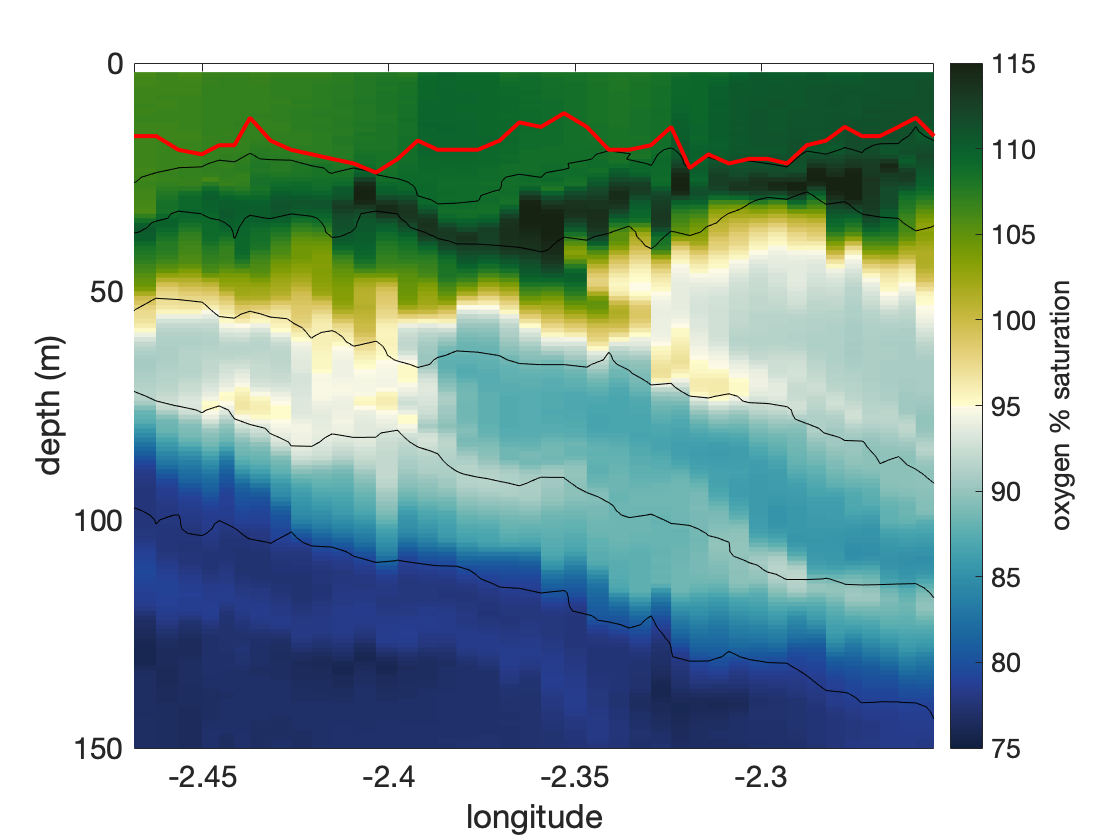}
    \caption{Depth structure of the northern transect on the V-shaped survey. The black lines are density contours in kg m$^{-3}$. Oxygen is plotted as percent saturation. Depth is in meters. The red line shows the mixed layer depth. Observations collected in collaboration with Mathieu Dever.}
    \label{fig:transect}
\end{figure}

Throughout the research cruise the depth structure of the hydrography and biogeochemistry were continuously being surveyed with an EcoCTD \cite{dever2020ecoctd} and vessel-mounted ADCP. In the location where water samples were taken, the surface is stratified and has nearly uniform density across the section while the density contours slope downwards forming a front subsurface (Fig. \ref{fig:transect}). There are intrusions of high oxygen water as a result of vertical motion. 

\begin{figure}
    \centering
    \includegraphics[width = 0.8\textwidth]{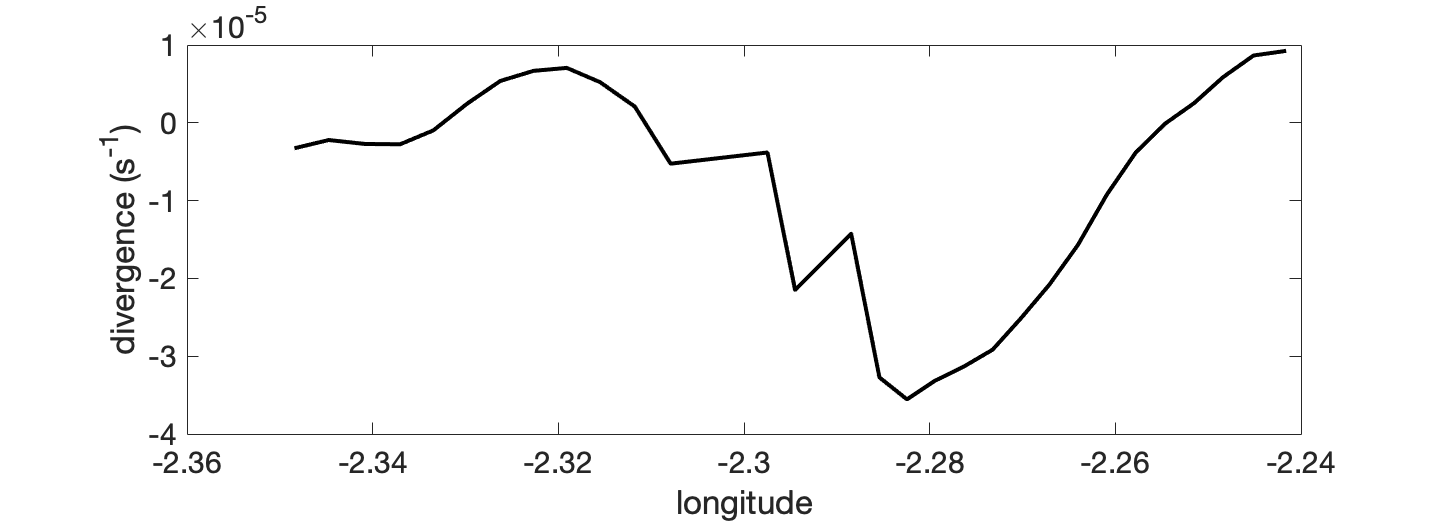}
    \caption{\new{Divergence of the along-track velocity on the southern observational transect.}}
    \label{fig:div_obs}
\end{figure}

\new{There is a noticeable velocity convergence on the upstream (southern) leg of the observational transect (Fig. \ref{fig:div_obs}).}

\subsection{Identification of cyanobacteria populations by flow cytometry}
\begin{figure}
    \centering
    \includegraphics[width = \textwidth]{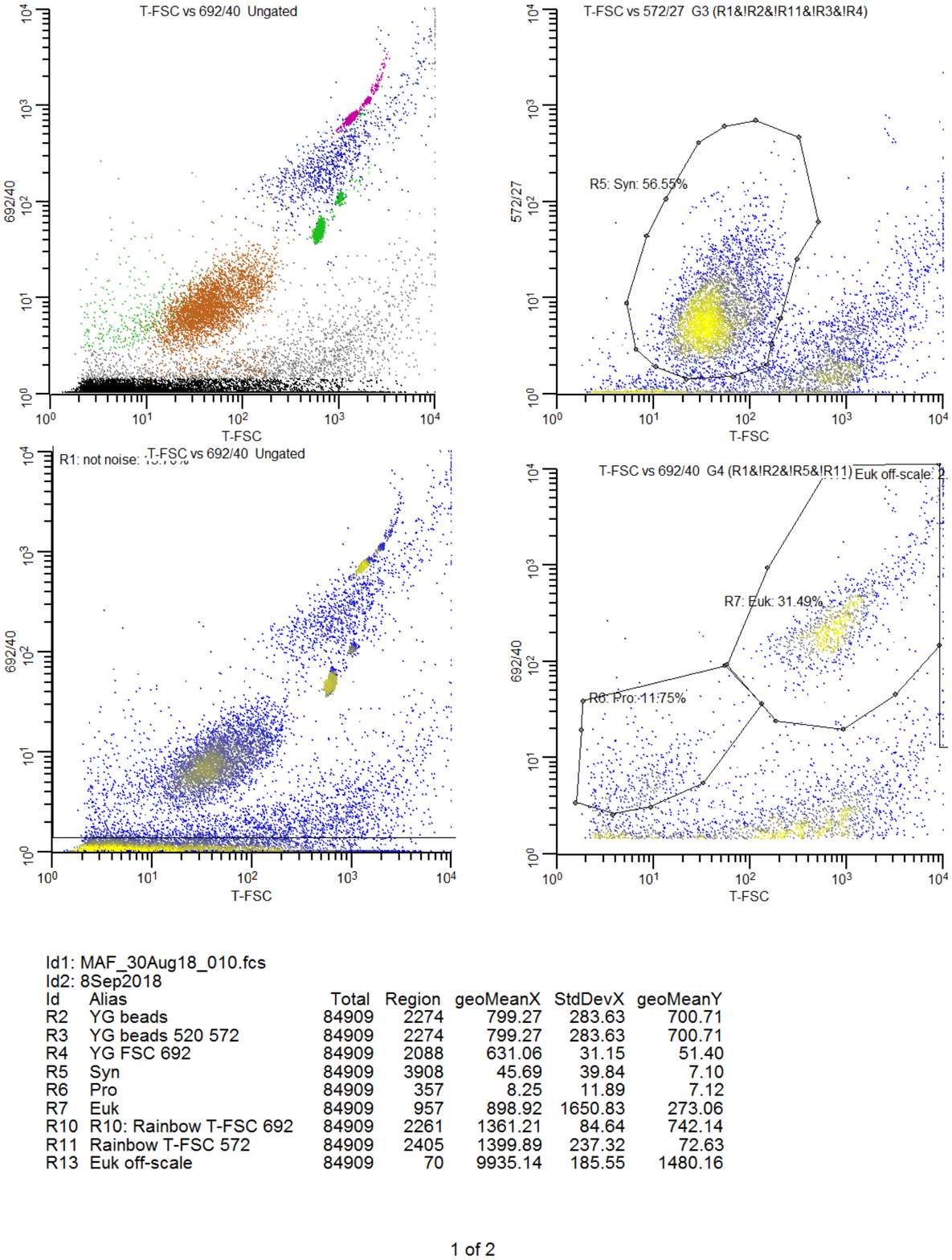}
    \caption{Winlist files showing identification of populations in flow cytometry. Top left: identified populations are identified in colors. Bright green and pink are YG and rainbow beads, respectively. Blue is Eukaryotes, orange is {\em Synechococcus}, green is {\em Prochlorococcus}. Bottom left: Ungated populations with color identifying density of observations. Top right: gating of {\em Synechococcus}. Bottom right: gating of {\em Prochlorococcus} and Eukaryotes}
    \label{fig:gates}
\end{figure}

The flow cytometry output files were manually analyzed using the WinList software. {\em Synechococcus} was identified using 572 autofluorescence and FALS (FSC). An example is shown in figure \ref{fig:gates}.

\subsection{Distribution of  \textit{Synechococcus}}
\begin{figure}
    \centering
    \includegraphics[width = 0.75\textwidth]{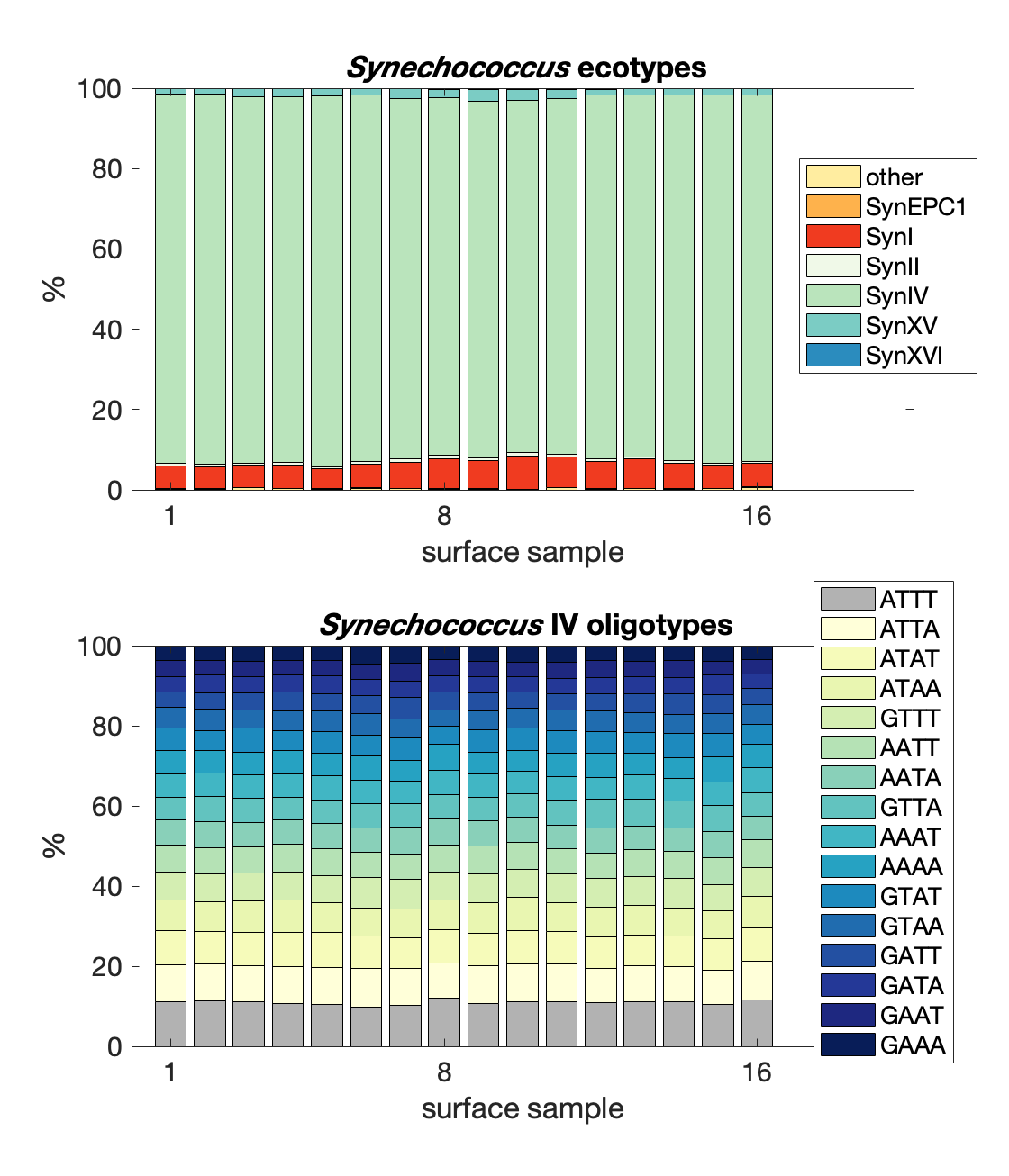}
    \caption{Community composition of bucket samples as relative abundance times 100, i.e. percent abundance. Top panel shows {\em Synechococcus} ecotypes and the lower panel shows oligotypes of the {\em Synechococcus} IV ecotype.}
    \label{fig:composition}
\end{figure}

\begin{figure}
    \centering
    \includegraphics[width = 0.6\textwidth]{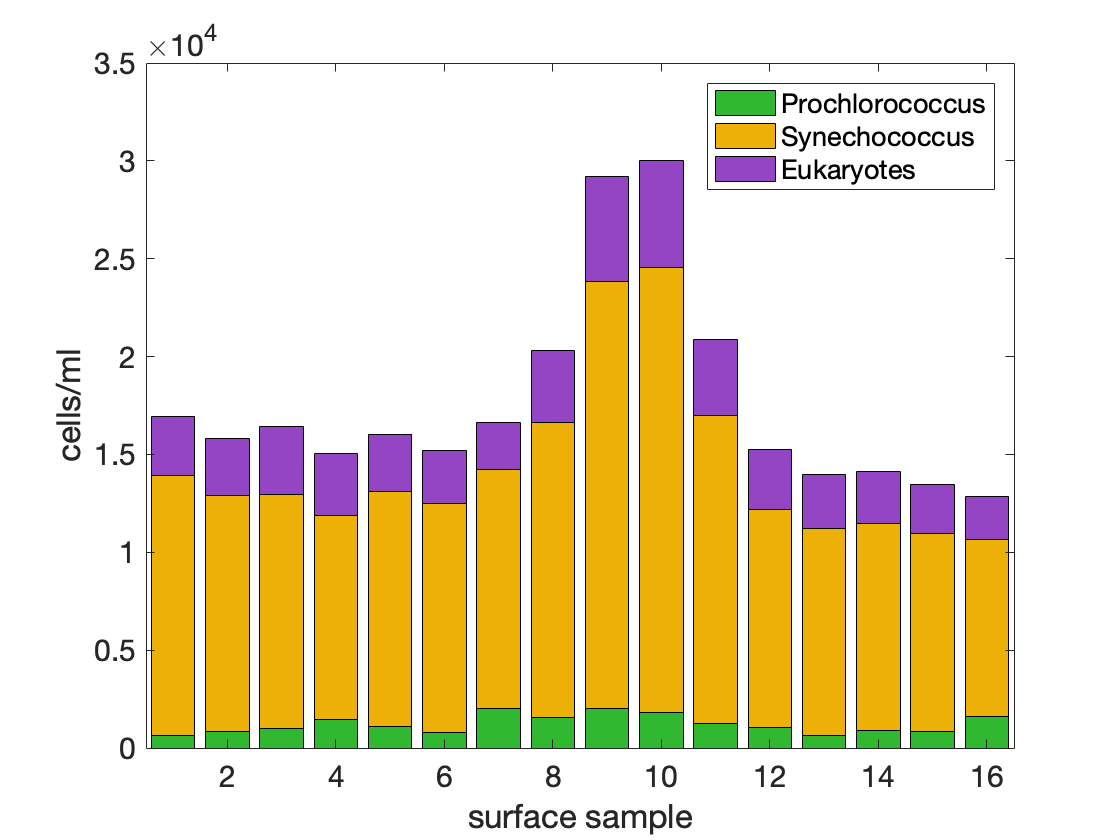}
    \caption{Community composition of bucket samples. Bars show cells per ml of \textit{Prochlorococcus}, \textit{Synechococcus}, and picoeukaryotes enumerated by flow cytometery.}
    \label{fig:counts}
\end{figure}

\begin{figure}
    \centering
    \includegraphics[width = 0.6\textwidth]{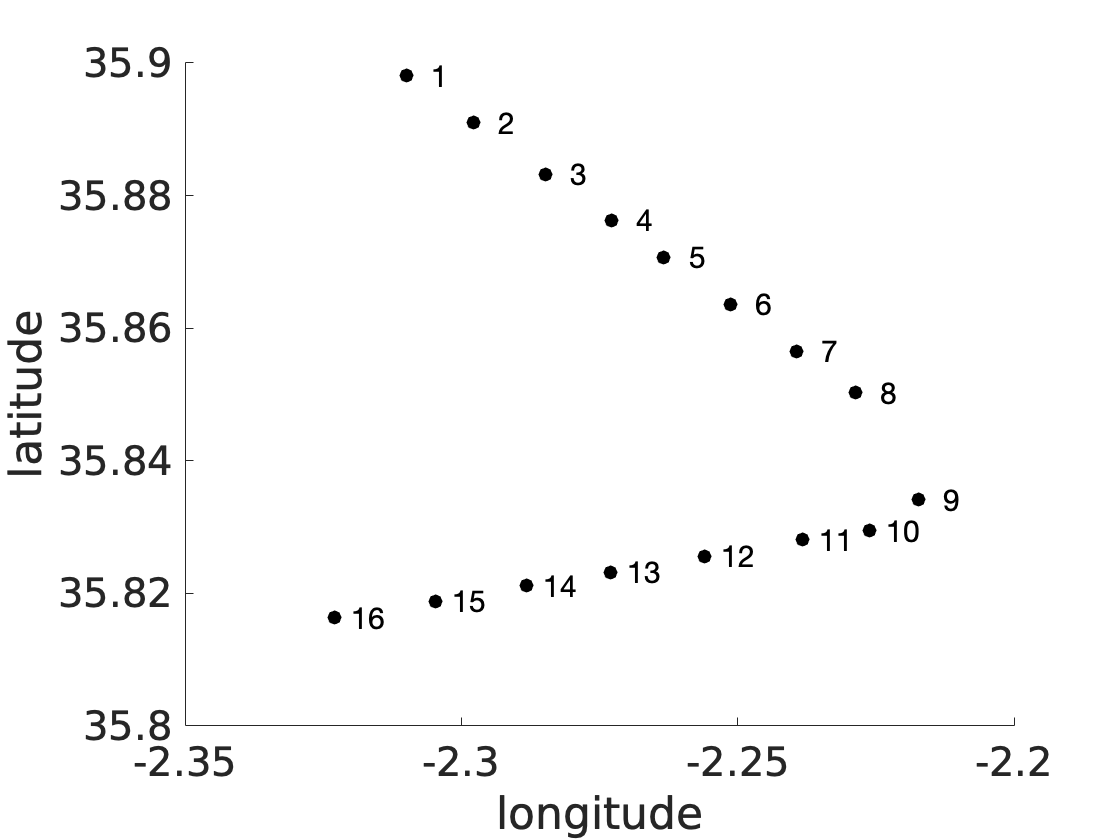}
    \caption{Locations of surface samples in Fig. \ref{fig:composition}.}
    \label{fig:sample_locations}
\end{figure}

\new{The phytoplankton community is dominated by a single ecotype of {\em Synechococcus}, clade IV. {\em Synechococcus} clade IV consistutes 89--92\% of the {\em Synechococcus} sequences (Fig. \ref{fig:composition}). Previous observations suggest that {\em Synechococcus} IV is commonly the dominant {\em Synechococcus} in surface waters in the Albor{\'a}n Sea \cite{mella2011distribution}. Quantitatively, {\em Synechococcus} is the dominant biological population in these surface samples with 9,080--22,700 cells ml$^{-1}$ compared with just 630--2,000 cells ml$^{-1}$ {\em Prochlorococcus} and 2,200--5,500 cells ml$^{-1}$ eukaryotes (Fig. \ref{fig:counts}). Although its abundance is relatively constant, the two easternmost samples have noticeably higher {\em Synechococcus} abundance (Fig. \ref{fig:counts}). There are 16~oligotypes of {\em Synechococcus} clade IV and although they are fairly evenly distributed, there is one oligotype that has the highest relative abundance across all samples, the ATTT oligotype (Fig. \ref{fig:composition}). Fig. \ref{fig:sample_locations} is provided for interpretation of the geographic locations of the samples in the two previous figures. While only the surface was sampled on this particular transect, in other samples taken on the same research cruise in nearby locations, {\em Synechococcus} populations are at highest abundance in the upper 50~m of the water column and decrease in abundance below that depth.} 

\new{In other oceanographic observations, dispersal and lateral mixing by mesoscale ocean currents has been shown to play an important role in shaping the distribution of phytoplankton populations \cite{d2010fluid}. To explore the influence of dispersal, and in particular of interleaving of nearby populations, on the spatial distribution of the observed populations, we trace the origin of the sampled populations backwards in time for 10~days using geostrophic velocity fields estimated from satellite altimetry (Fig. \ref{fig:backwards}). These velocity fields have 1/4$^\circ$ resolution. The sampled high chlorophyll filament appears to have originated 10~days earlier near the Spanish coast in a region with frequent upwelling \cite{sarhan2000upwelling}. The water masses were then advected by the Western Albor{\'a}n Gyre and reached the sampled location on the western edge of the Eastern Albor{\'a}n Gyre. Although there is some evidence of geostrophic straining with the populations converging on the edge of the Western Albor{\'a}n Gyre, there is not evidence of interleaving. Instead, the water masses appear to have followed nearly parallel trajectories through the flow field. Interleaving could occur due to currents that are below the resolution of the altimeter or as a consequence to divergent currents that are not included in estimates of velocity from altimetry.}

\begin{figure}
    \centering
    \includegraphics[width = 0.5\textwidth]{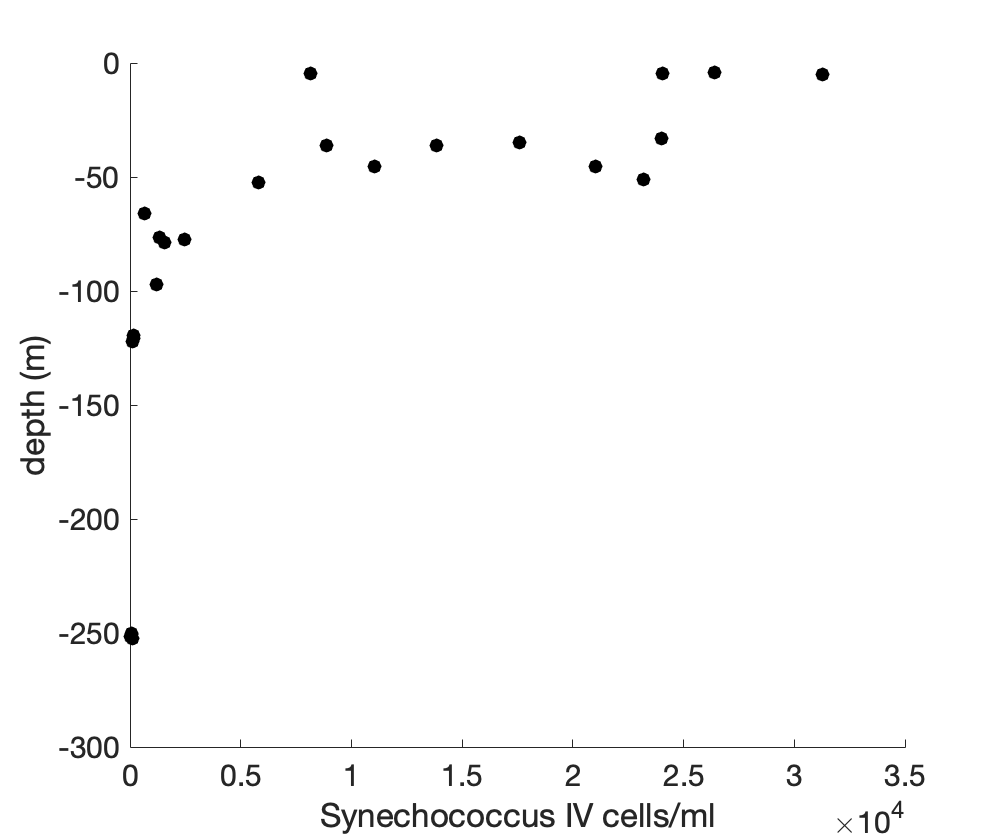}
    \caption{Vertical profile of \textit{Synechoccocus} IV near the sampled region}
    \label{fig:SynIV}
\end{figure}

\begin{figure}
    \centering
    \includegraphics[width = \textwidth]{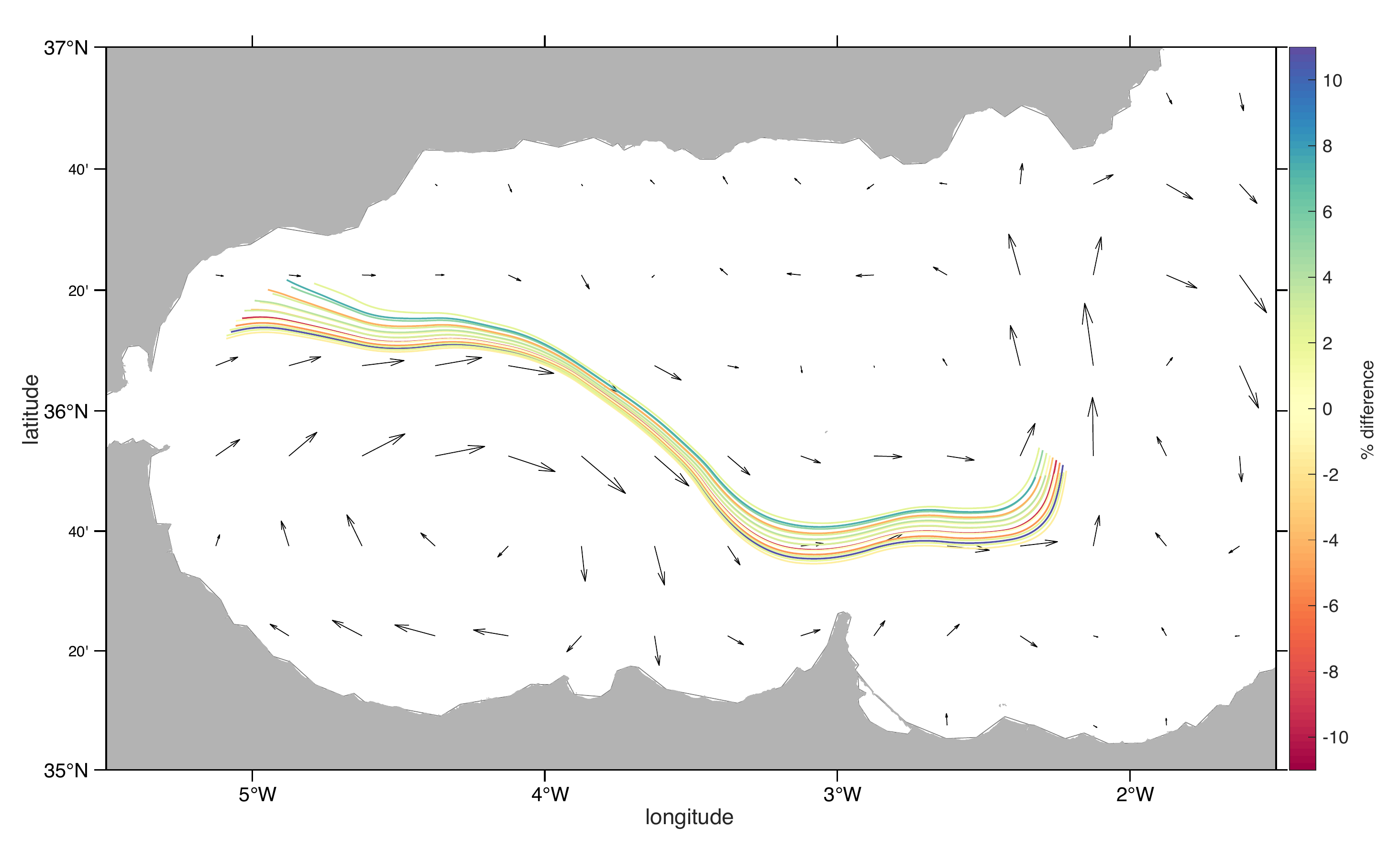}
    \caption{\new{Stirring of biological communities. Each line traces the path of the sampled water parcels backwards in time using the satellite geostrophic velocity. The line color indicates the difference in the relative abundance of oligotype ATTT when compared with the population in sample 12 (sample numbers labeled in figure \ref{fig:sample_locations}).}}
    \label{fig:backwards}
\end{figure}

\subsection{Constant depth approximation}
In the main text, we simulate competition on constant depth fluid surfaces. 

We could have instead modeled competition on constant density, or isopycnal, surfaces. We would not expect this to weaken the impact of effective compressibility on the ecology, since the distribution of divergence on relevant isopycnal surfaces is similar to the distribution of divergence on the corresponding constant depth surfaces, as we discuss below. However, depth variations of isopycnal surfaces also lead to variations in light, which affect the carrying capacity of the fluid. Simulating the biological model on constant depth surfaces allows us to better isolate the effects of divergence. 

\begin{figure}
\begin{center}
\includegraphics[width=0.9\textwidth]{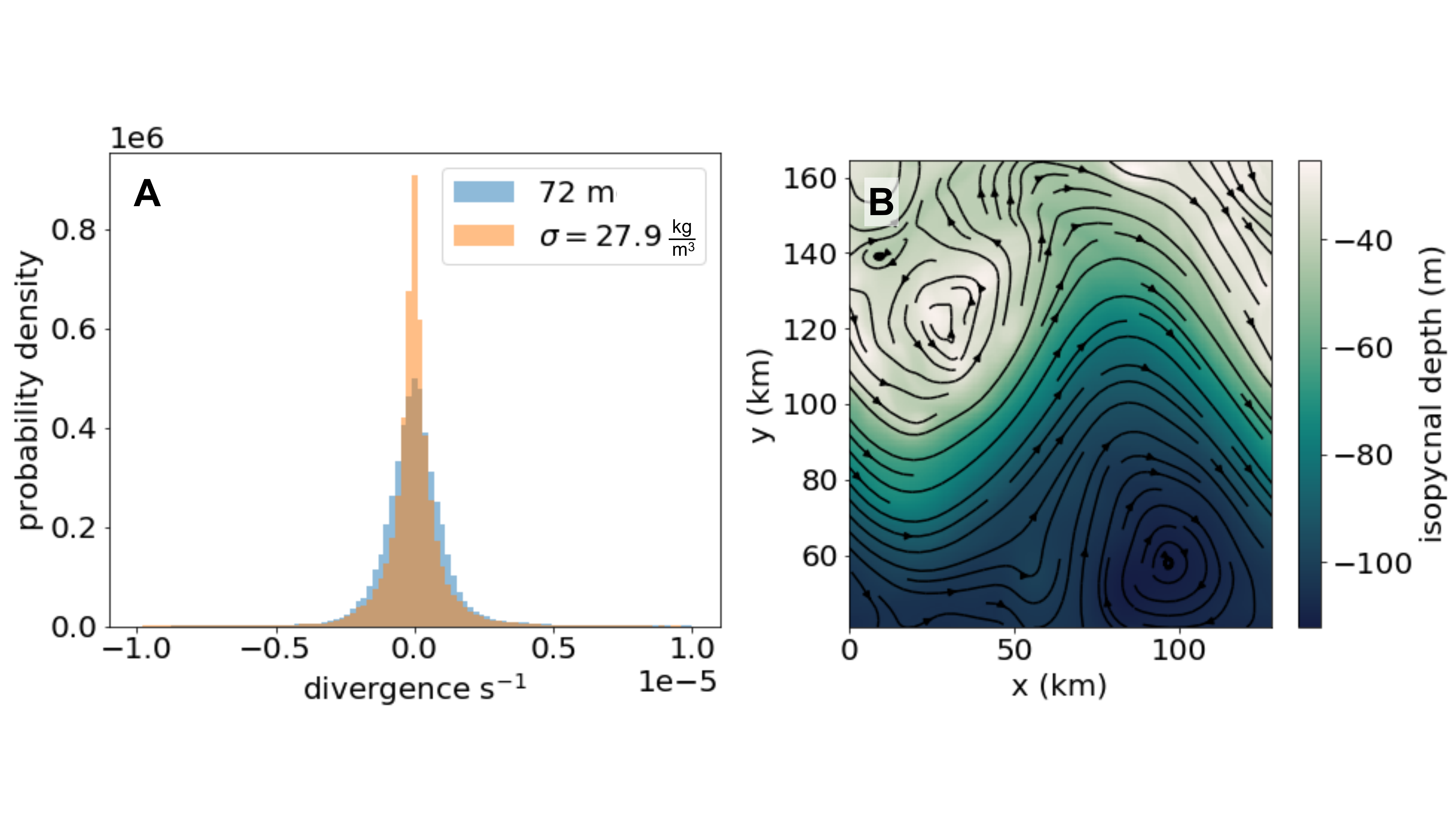}
\caption{A. Distribution of divergence at a fixed depth of 72 m (blue) and on an isopycnal surface corresponding to a density of $\sigma=27.9 \text{ kg}\text{ m}^{-3}$ with a mean depth of 72 m (orange). Divergence is evaluated at the grid scale, 500 m, as in Fig. \ref{fig:hov_div}. B. The isopycnal surface $\sigma=27.9 \text{ kg}\text{ m}^{-3}$, with the directions of velocities tangent to the surface given with the arrows. {\label{isopycnals}
}}
\end{center}
\end{figure}

We present data for the flows tangent to isopycnal and constant depth surfaces in Fig. \ref{isopycnals}A. We compare the distribution of divergence on an isopycnal surface from the winter flow field at a density $\sigma=27.9 \text{ kg} \text{ m}^{-3}$, pictured in Fig. \ref{isopycnals}B, to the divergence on a fixed depth surface at the average depth of that isopycnal surface. \new{Divergence on an isopycnal surface is calculated by first transforming the 3-dimensional velocity vector from depth to isopycnal coordinates then by computing the (2-dimensional) divergence of the flow the isopycnal surface}. We see that the distributions are similar.

The isopycnal surface used for comparison in Fig. \ref{isopycnals} is on average deeper than both the winter flow field constant depth surface (52 m) and the summer flow field constant depth surface (0 m) used in the main text. We do not show isopycnal surfaces with these average depths because they outcrop at the sea surface. Instead, we note that isopycnals with shallower average depths in these models also span a smaller depth range and in general have even stronger regions of divergence due to the larger divergence at the sea surface.

\subsection{Weak compressibility regime} \label{weakcomp}
The strength of the effective compressibility experienced by a population depends on both the magnitude of the divergence and the growth characteristics of the population, since rapidly reproducing phytoplankton can maintain a constant density even in the presence of a positively divergent flow \cite{perlekar2013cumulative}. 

In the main text, we often assume that we are in the weak compressibility regime, and the steady state concentration profile is close to the concentration profile in the absence of flow ($c(\*x) \approx 1$). This assumption simplifies the analysis considerably, and also makes the uniformly occupied domains used as initial conditions in the simulations more reasonable (strong compressibility would lead to localization on downwellings).  

The weak compressibility regime was also the focus of Ref. \cite{plummer}, where it was defined to mean that Fisher population waves are able to propagate through regions of convergence without becoming trapped. We can estimate whether this condition holds using the data provided in Fig. \ref{fig:hov_div}. At the grid scale of the fluid model, 500 m, the strongest regions of divergence in both winter and summer flow fields measure approximately $4 \times 10^{-6} \text{ s}^{-1}$. Upon comparing to the Fisher velocity for $\mu=1 \text{ day}^{-1}$, we find
\begin{equation}
    (500 \text{ m})(4 \times 10^{-6} \text{ s}^{-1})= 0.002\text{ m}/\text{s}< 2 \sqrt{D \mu}\approx 0.007\text{ m}/\text{s}.
\end{equation}
The inequality is even stronger for the $\mu= 2 \text{ day}^{-1}$ and $D=5 \text{ m}^2/\text{s}$ simulations.

We can also directly compare the time scale of the strongest divergence on the grid scale to the time scale of replication to find a dimensionless measure of flow divergence. For $\mu= 1 \text{ day}^{-1}$,
\begin{equation}
    \frac{|\nabla \cdot \*u|}{\mu}\approx 0.35.
\end{equation}

The generation time, $\mu^{-1}$, is therefore short relative to the source time. Organisms are able to reproduce multiple times while feeling the influence of even the strongest sources in the flow, if the sources are traveling with the mean flow, before being moved by the source itself away from the area of interest.

\begin{figure}
\begin{center}
\includegraphics[width=0.45\textwidth]{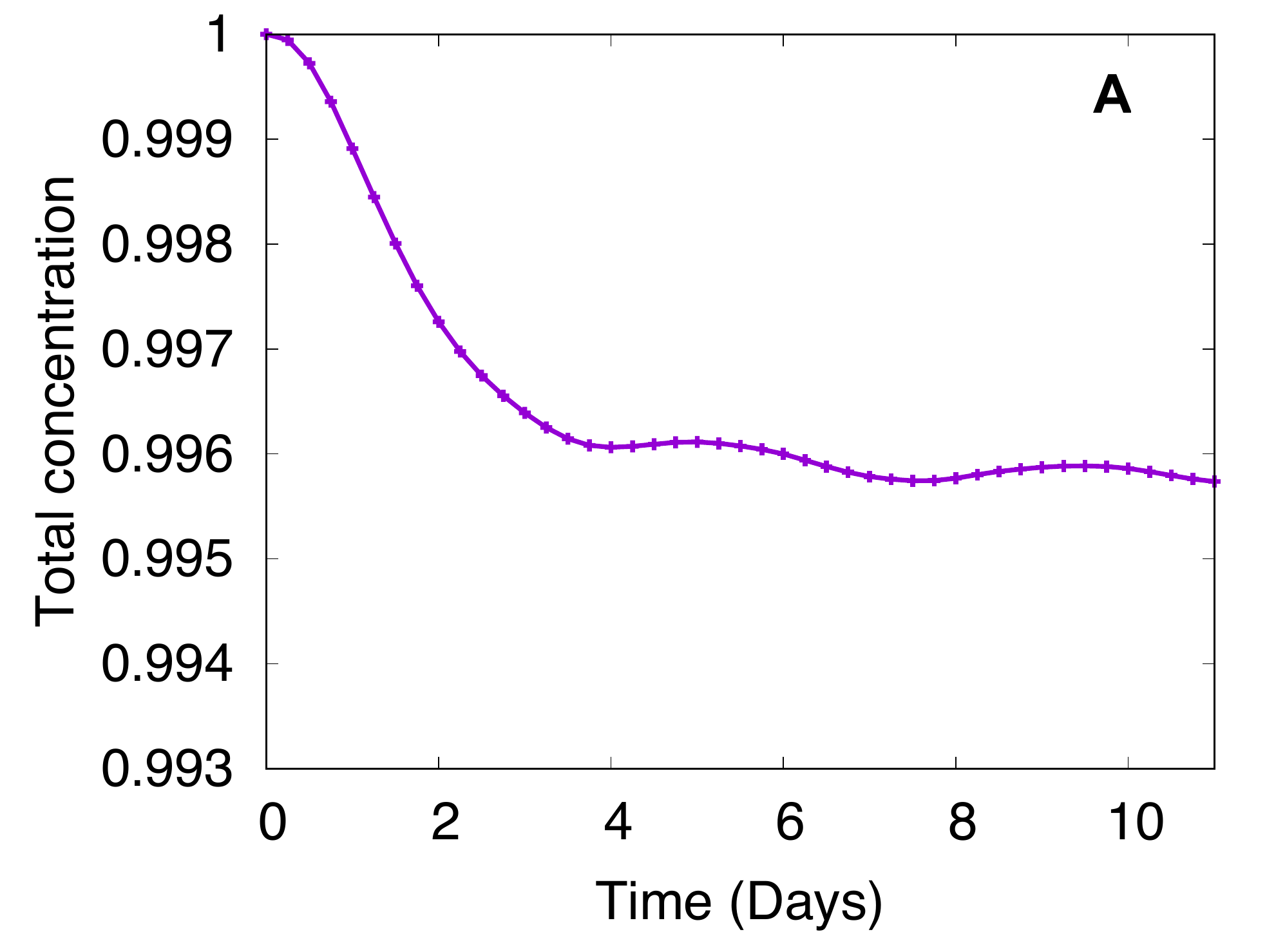}
\includegraphics[width=0.45\textwidth]{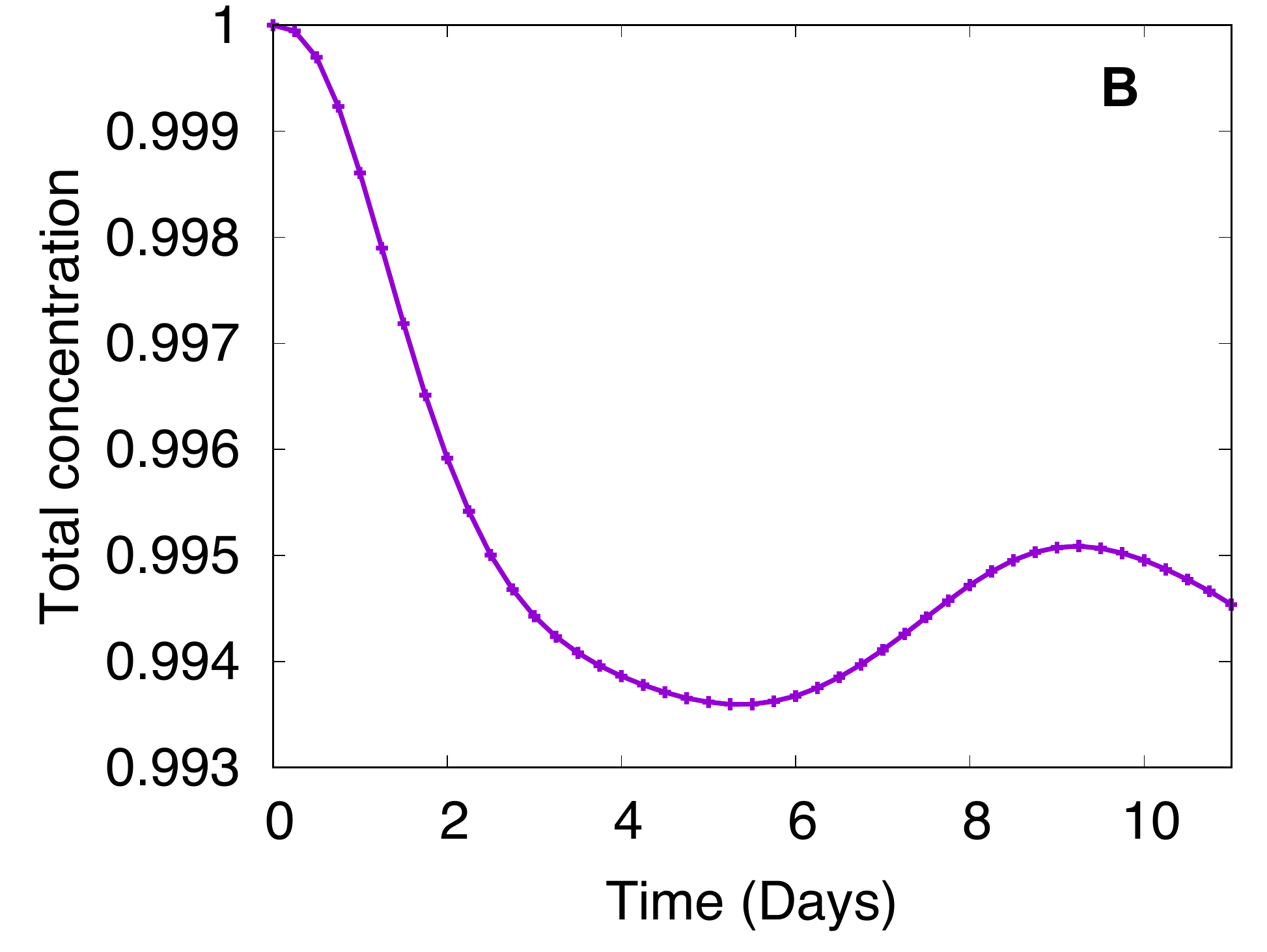}
\end{center}
\caption{{\label{longrun} Total concentration (relative to the no-flow carrying capacity) as a function of time for the (A) winter flow field and (B) summer flow field, with $\mu= 1 \text{ day}^{-1}$.
}}
\end{figure}

These estimates, while useful, do not consider any effects resulting from sources and sinks moving relative to the mean flow. To be certain that the flow does not induce a significant reduction in the carrying capacity, we measure the total concentration over ten days of the simulation. If Fisher population waves can overcome the convergences, the system will remain approximately uniformly occupied. As we see in Fig. \ref{longrun}, the total concentration remains approximately equal to the concentration in the absence of flow (normalized to 1 in the figure) as the velocity field changes over time.

Together, these arguments suggest that for both the summer and winter flows, we are comfortably in the weak compressibility regime. If we had instead found strong compressibility, we would still expect to observe profound effects on competition events. However, for strong compressibility, the population tends to localize on sinks \cite{benzi2009fisher}, which makes the effect of regions of positive divergence more subtle, with number fluctuations becoming important. This regime would be interesting to study in the future. 

\subsection{Relative abundance in incompressible and compressible flows}
We begin with the evolution equations introduced in the main text. 
\begin{align}
    \frac{\partial c_{A}}{\partial t}+\nabla \cdot\left(\mathbf{u} c_{A}\right)&=D \nabla^{2} c_{A}+\mu c_{A}\left(1-c_{A}-c_{B}\right)+s \mu  c_{A} c_{B},\label{ca}\\
    \frac{\partial c_{B}}{\partial t}+\nabla \cdot\left(\mathbf{u} c_{B}\right)&=D \nabla^{2} c_{B}+\mu c_{B}\left(1-c_{A}-c_{B}\right)-s \mu c_{A} c_{B}. \label{cb}
\end{align}

We are interested in how the fraction of population $A$ changes in response to the external flow field. Therefore, we change variables and rewrite these equations in terms of $f= \frac{c_A}{c_A+c_B}$ and $c=c_A+c_B$. We do this with the following steps. 

First consider the sum of Eqs. \ref{ca} and \ref{cb}. The term concerning selection cancels (although it would not if we had assumed the selective advantage occurs due to a difference in growth rates, as we discuss in SI Sec. \ref{selection}).

\begin{equation}
 \frac{\partial c}{\partial t} + \nabla \cdot  (\*u c) = D \nabla^2 c + \mu c (1- c).
    \label{cfinal}
\end{equation}

This equation gives the evolution of the total concentration field. To find the equation for the relative abundance, we take Eq. \ref{ca} and substitute $c_A=f c$ and $c_B = c-f c$. The time derivative on the left hand side of Eq. \ref{ca} becomes

\begin{equation}
    \frac{\partial (f c)}{\partial t}= c \frac{\partial f}{\partial t} + f \frac{\partial c}{\partial t} =  c \frac{\partial f}{\partial t} + f \left(-\nabla \cdot  (\*u c)+ D \nabla^2 c + \mu c (1- c) \right) . 
\end{equation}
We simplify, distributing the divergence operator, to find
\begin{equation}
\frac{\partial f}{\partial t} + \*u \cdot \nabla f = D \nabla^2 f + \frac{2 D}{c} \nabla f \cdot \nabla c + s \mu  c f(1 - f).
    \label{fracfinal}
\end{equation}

Next, we consider the specific scenario treated in the main text. When the total concentration is initially uniform in space, it will remain uniform for incompressible flows and approximately uniform for weakly compressible flows (SI Sec. \ref{weakcomp}).
Therefore, we neglect the $\frac{2 D}{c} \nabla f \cdot \nabla c$ term and set $c=1$ in the selection term. 

We integrate the remaining terms over the domain, integrate by parts, and apply Gauss's theorem. 
\begin{equation}
\frac{\partial}{\partial t} \int_\Omega  f d \Omega\approx D \oint_S  \nabla f 
\cdot \hat{\*n}dS-\oint_S f \*u \cdot \hat{\*n}dS  + \int_\Omega  f (\nabla \cdot \*u)d\Omega+ s \mu \int_\Omega  f(1-f) d\Omega.
    \label{frac3}
\end{equation}
The surface terms are zero for a sufficiently localized population of $A$ ($\nabla f=f=0$ at the boundary), leaving the condition
\begin{equation}
\frac{\partial}{\partial t} \int_\Omega f d \Omega\approx \int_\Omega  f (\nabla \cdot \*u)d\Omega +s \mu \int_A f(1-f) d\Omega.
    \label{frac4}
\end{equation}
Therefore, for an incompressible flow with $\nabla \cdot \*u=0$, the spatially averaged relative abundances remain constant in time for an initial condition with a localized population of type $A$ embedded in a community of neutral competitors of type $B$ such that $c=c_A+c_B$ is uniform in space. \new{Furthermore, if $c=1$ initially, $c=1$ for all time and $f=c_A$.} We can therefore attribute the changes we see in Figs. 2 and 3 of the main text to the influence of divergence. 

 We note that this argument can also be applied for a carrying capacity that varies sufficiently slowly in space such that the $\frac{2 D}{c} \nabla f \cdot \nabla c$ term can be neglected. \neww{Dimensional analysis shows that this simplification is justified in the submesoscale dynamical regime. Assuming weak compressibility, the total concentration will locally attain its local carrying capacity on average (defined relative to a reference patch with carrying capacity rescaled to 1). Given length scale of variations in relative abundance $l$, length scale of variations of the carrying capacity/concentration $L$, velocity scale $U$, diffusivity scale $d$, relative abundance magnitude $F$, and total concentration magnitude $C$, we can write a non-dimensional expression comparing terms in the evolution equation for relative abundance. We compare the advection term to the term with gradients in the concentration. We denote non-dimensional variables as $\hat{\cdot}$.}
\begin{equation}
    \neww{\frac{d}{C}\frac{F}{l}\frac{C}{L}\frac{2 \hat{D}}{\hat{c}} \hat{\nabla} \hat{f} \cdot \hat{\nabla} \hat{c} \sim U\frac{F}{l}\hat{\*u} \cdot \hat{\nabla} \hat{f}}
\end{equation}
\neww{The relative magnitude of these two terms can be assessed through the non-dimensional parameter $\frac{d}{LU}$. The terms are comparable if $\frac{d}{LU} \sim 1$. Given typical parameter values in the submesoscale regime, including in the models used in this study, of $d \sim \mathcal{O}(1~\rm{m s}^{-1})$ and $U \sim \mathcal{O}(0.1~\rm{m s}^{-1})$, the terms are comparable if $L \sim \mathcal{O}(10~\rm{m})$. However, the submesoscale regime considers scales of $\mathcal{O}(1-10~\rm{km})$ in both velocity and tracers (and the model grid scale is 500~m) so we can conclude that the term $\frac{2 D}{c} \nabla f \cdot \nabla c$ can be neglected.}

\subsection{Selective advantage}\label{selection}

Here, we consider the influence of selection using the governing equations, relevant to the simulation results presented in Figs. 4 and 5 of the main text. 

With selection, Eq. 10 of the main text gains an additional term,
\begin{equation}
    \frac{\Delta \langle f \rangle }{f_0}\approx \frac{\tau}{f_0} \left(\overline{ \left\langle f \nabla \cdot \*u \right \rangle} + s \mu \overline{\left \langle  f (1-f) \right \rangle} \right).
    \label{fracchange}
\end{equation}
For biologically realistic selection between closely related populations, $s$ is small, making the characteristic time on which selection operates $(s \mu)^{-1}$ longer than one generation, $ \mu^{-1}$. Therefore, over our standard observation period $\tau=$1 day$=\mu^{-1}$, the difference between $f(\*x, \tau)$ for $s=0$ and $f(\*x, \tau)$ for $s \neq 0$ is small, and we expect $\Delta \langle f \rangle /f_0$ to be linear in $s$.  In the simulations using nonzero selective advantage (Figs. 4 and 5 of the main text), we use the same flow field and initial condition for each set of trials, only varying $s$, and we observe linear trends in Figs. 4 and 5, as expected. For sufficiently large $|s|$ or at long times, the data deviate from the linear trend. 

While Eq. \ref{fracchange} is useful for testing the validity of our approximations and understanding how regions of divergence interact with community structure in general, as we discuss in the next two sections, we gain a more intuitive understanding by considering further approximations for a simplified case.  

In the main text, we use a Gaussian initial condition with the fraction of organisms in the localized population given by 
\begin{equation}
    f(x,y)=\exp\left(-\frac{x^2+y^2}{2 \sigma^2} \right).
    \label{eq:gaussian}
\end{equation}
We now assume that this is the population structure for all time. This approximation is reasonable when $\tau$ is small relative to the time scales of advection and selection. 

We also assume that the localized population is centered on a source or sink of the velocity field, such that
\begin{equation}
    \*u(x,y)=\frac{\delta}{2}(\*x + \*y).
\end{equation}
We can think of this as the first term in a Taylor series expansion, accurate sufficiently close to a source or sink. If $\delta>0$, this is a source (positive divergence). If $\delta<0$, this is a sink (negative divergence). This approximation is especially good for the simulations discussed in the main text where we place the localized population on a region of high positive divergence. 

Integrating over all space (extending the bounds to $\pm \infty$ for simplicity), we find
\begin{equation}
    \frac{\Delta \langle f\rangle}{f_0}\approx \tau \left(\delta+\frac{\mu s}{2} \right).\label{Gaussian}
\end{equation}
This result has a number of features that agree with our simulations. As we saw in Fig. 3 of the main text, the change in the relative abundance is linearly proportional to the strength of the divergence, and crosses zero at $\delta=0$ when $s=0$. We do not have a dependence on the diffusivity, the localized population size, or the growth rate when $s=0$. 

This argument also predicts that the change in relative abundance should be linearly proportional to the selective advantage with a slope of 0.50 for $\mu= 1 \text{ day}^{-1}$, and that the change in relative abundance will be zero when $s=s^*$, with
\begin{equation}
    s^*=-\frac{2 \delta}{\mu}.
    \label{sstar}
\end{equation}

We can compare these theoretical expectations to the results presented by Fig. 4 of the main text. We find good agreement for the predicted slope: a linear trend for both trials with a slope of 0.53 (R$^2$ = 0.99) for the summer flow field and a slope of 0.47 (R$^2$ = 0.99) for the winter flow field. The winter flow field initial condition is centered on a region of divergence with $\delta=1.25 \times 10^{-6} \text{ s}^{-1}$ (Fig. 4 A). Substituting this value into Eq. \ref{sstar}, we estimate $s^*\approx -0.22$. The measured value of $s^*$ is $-0.25$. Agreement is somewhat worse for the summer flow field initial condition centered on a region of divergence with $\delta=2.99 \times 10^{-6} \text{ s}^{-1}$. Eq. \ref{sstar} gives an estimate of $s^*\approx -0.52$, while the measured $s^*$ is $-0.65$. Considering the severity of the approximations leading to Eq. \ref{sstar}, these estimates are perhaps surprisingly close to the measured values, which take into account nonlinearities and the time-dependent nature of the velocity field.
 
\subsubsection*{Other forms of selective advantage}
As described in many places \cite{risken, gardiner, pigolotti2012population, plummer}, macroscopic equations for the concentration fields can be derived from microscopic, agent-based rules. The form of the selective advantage term in the (macroscopic) evolution equations used in this work (Eqs. \ref{ca} and \ref{cb}) arises from reducing the microscopic rate of death-by-competition for population $A$, thereby giving population $A$ an advantage. 

However, selective advantage is often instead modeled as a difference in growth rates. Coarse-graining with two different growth rates, $\mu$ and $\mu(1+s)$, results in different evolution equations.

Eq. \ref{cfinal} becomes
\begin{equation}
    \frac{\partial c}{\partial t} + \nabla \cdot  (\*u c) = D \nabla^2 c + \mu c (1+fs - c).
\end{equation}
Eq. \ref{fracfinal} becomes
\begin{equation}
    \frac{\partial f}{\partial t} + \*u \cdot \nabla f = D \nabla^2 f + \frac{2 D}{c} \nabla f \cdot \nabla c + s \mu  f(1 - f).
\end{equation}

Note that there is now a term in the equation for $c$ that depends on $s$. Since type $A$ has a faster growth rate, its equilibrium carrying capacity in isolation is greater than that of type $B$. Therefore, the total carrying capacity of the system now depends on the fraction $f$. Assuming weak compressibility, we set $c=1+s f$. 

If we make the same Gaussian population/linear source approximations as in the previous section, we find, to first order in $s$,
\begin{equation}
    \frac{\partial}{\partial t}\int_\Omega f d\Omega \approx \left(2 \delta+s \mu \right) \pi \sigma^2 + 2 D s \pi,
\end{equation}
\begin{equation}
    \frac{\Delta \langle f\rangle }{f_0} \approx \tau\left(\delta+\frac{\mu s}{2} + \frac{D s}{\sigma^2}\right) .
\end{equation}

Since $\frac{D}{\sigma^2}< \frac{\mu}{2}$ (the amount of time it takes an organism to diffuse across the localized population is longer than two generations) in simulations, the correction due to birth-based selection is generally small.

\bibliographystyle{unsrtnat}
